\documentclass[12pt]{article}

\usepackage[english]{babel}

\usepackage[letterpaper,top=2cm,bottom=2cm,left=3cm,right=3cm,marginparwidth=1.75cm]{geometry}

\usepackage{graphicx}
\usepackage{amssymb}
\usepackage[T1]{fontenc}
\usepackage{booktabs, multirow} 
\usepackage[colorlinks=true, allcolors= black]{hyperref}
\usepackage{soul}
\usepackage[table]{xcolor} 
\usepackage{changepage,threeparttable} 
\usepackage{lscape} 
\usepackage{longtable}
\usepackage{natbib}
\usepackage{adjustbox}
\usepackage[nottoc]{tocbibind}
\usepackage{authblk}
\DeclareUnicodeCharacter{2212}{-}
\DeclareUnicodeCharacter{03C3}{}
\usepackage[shortcuts]{extdash}

\usepackage{setspace}
\usepackage{gensymb}
 
\date{} 
\title{Lifetime of the Outer Solar System Nebula From Carbonaceous Chondrites}

\author[1,2]{Cauê S. Borlina*}
\author[2]{Benjamin P. Weiss}
\author[3]{James F. J. Bryson}
\author[4,5]{Philip J. Armitage}

\affil[1]{Department of Earth and Planetary Sciences, Johns Hopkins University, Baltimore, MD, USA}
\affil[2]{Department of Earth, Atmospheric and Planetary Sciences, Massachusetts Institute of Technology, Cambridge, MA, USA}
\affil[3]{Department of Earth Sciences, Oxford University, Oxford, UK}
\affil[4]{Center for Computational Astrophysics, Flatiron Institute, New York, NY, USA}
\affil[5]{Department of Physics and Astronomy, Stony Brook University, Stony Brook, NY, USA}

\begin{document}
\maketitle

\begin{center}
\normalsize{$^\ast$Corresponding author. Email: \href{mailto:csciasc1@jhu.edu}{csciasc1@jhu.edu}}\\
\end{center}

\begin{center}
See most updated version published in \textit{JGR: Planets} \href{https://doi.org/10.1029/2021JE007139}{https://doi.org/10.1029/2021JE007139}\\
\end{center}

\begin{abstract}
The evolution and lifetime of protoplanetary disks (PPDs) play a central role in the formation and architecture of planetary systems. Astronomical observations suggest that PPDs go through a two-timescale evolution, accreting onto the star over a few to several million years (Ma) followed by gas-dissipation within $\lesssim$1 Ma. This timeline is consistent with gas dissipation by photoevaporation and/or magnetohydrodynamic winds. Because solar nebula magnetic fields are sustained by the gas of the protoplanetary disk, we can use paleomagnetic measurements to infer the lifetime of the disk. Here we use paleomagnetic measurements of meteorites to investigate whether the disk that formed our solar system had a two-timescale evolution. We report on paleomagnetic measurements of bulk subsamples of two CO carbonaceous chondrites: Allan Hills A77307 and Dominion Range 08006. If magnetite in these meteorites could acquire a crystallization remanent magnetization that recorded the ambient field during aqueous alteration, our measurements suggest that the local magnetic field strength at the CO parent-body location was <0.9 µT at some time between 2.7-5.1 million years (Myr) after the formation of calcium-aluminum-rich inclusions. Coupled with previous paleomagnetic studies, we conclude that dissipation of the solar nebula in the 3-7 AU region occurred <1.5 Myr after the dissipation of the nebula in the 1-3 AU region, suggesting that protoplanetary disks go through a two-timescale evolution in their lifetime. We also discuss future directions necessary to obtain robust records of solar nebula fields using bulk chondrites, including obtaining ages from relevant chondrites and experimental work to determine how magnetite acquires magnetization during chondrite parent-body alteration.  
\end{abstract}

\section{Introduction}
The lifetimes of protoplanetary disks (PPDs) have important implications for the formation of planetary systems that emerge from them. Determining the lifetime of PPDs (i.e., the dissipation of the gas) constrains the formation time of gas giants and when gas-driven planetary migration can occur. The lifetimes of PPDs have been measured astronomically through observations of dust infrared spectral energy distributions (SEDs) \citep{haisch_disk_2001, hernandez_spitzer_2007}. SED observations have provided evidence that disks pass through different stages during their evolution from that of protostars to transition disks (disks that have low or no near-infrared excess and high mid- to far-infrared excess) and to debris disks (no remaining gas) \citep{ercolano_dispersal_2017, owen_origin_2016}. With the exception of a few observed PPDs that are likely in the transition phase, the majority (~90\%) of observed PPDs are either in PPD phase with a protostar or completely dissipated \citep{owen_origin_2016}. This observation has been interpreted to suggest that PPDs have a two time-scale evolution during their lifetime: during the first evolutionary stage, which lasts $\sim$ 3-5 million years (Myr), the disk evolves via magnetohydrodynamic and/or hydrodynamic processes, with angular momentum transported outwards that enables accretion onto the star \citep{audard_protoplanetary_2019, gorti_disk_2016}; during the second stage, the remaining gas disperses rapidly, in $\lesssim$ 1 Myr \citep{ercolano_dispersal_2017}. Several modes of disk dispersal have been proposed \citep{ercolano_dispersal_2017}, including photoevaporation by the central star \citep{alexander_photoevaporation_2006} or a nearby high-mass star \citep{concha-ramirez_external_2019, scally_destruction_2001}, magnetohydrodynamic (MHD) winds \citep{armitage_two_2013, bai_toward_2016, shadmehri_time-dependent_2019, suzuki_evolution_2016}, planet formation \citep{zhu_transitional_2011}, grain growth \citep{dullemond_dust_2005}and close-binary effects \citep{ireland_disk_2008}. The two-timescale evolution is generally consistent with gas dissipation by photoevaporation and/or magnetohydrodynamic winds. Because the lifetimes of PPDs are far greater than the times over which we can observe them, testing the two timescale hypothesis through observations of any one PPD is essentially impossible. 
However, the two-timescale hypothesis can be tested using paleomagnetic measurements of our solar system’s PPD, also known as the solar nebula. Due to the coupling of magnetic fields with gas, the nebular magnetic field is a proxy for the presence of gas in the disk, and thus the lifetime of the disk \citep{wang_lifetime_2017, weiss_history_2021}. By obtaining paleomagnetic measurements from meteorites that formed at different times and locations, it is possible to constrain the lifetime and spatial evolution of the PPD of our solar system. Previous paleomagnetic studies of meteorites have provided evidence for a large scale magnetic field present in the disk up to at least $3.9_{-0.5}^{+0.4}$ Myr after the formation of the calcium-aluminum-rich inclusions (CAIs) \citep{borlina_paleomagnetic_2021, cournede_early_2015, fu_solar_2014}. Additional studies have also found evidence for the dissipation of the gas at <3 AU (inner solar system) by 3.2-5.0 Myr after CAI formation, with Pb-Pb ages indicating a mean dissipation time by 3.71 $\pm$ 0.2 Myr after CAI-formation \citep{wang_lifetime_2017, weiss_history_2021}. 

Here we use carbonaceous chondrites to obtain better constraints on the timescale of the dissipation of the solar nebula. Carbonaceous chondrites are unmelted accretional aggregates of refractory inclusions (CAIs and amoeboid olivine aggregates), chondrules and matrix, which likely formed in the 3-7 AU region (i.e., outer solar system). In this study, we investigate the magnetic record of bulk subsamples of carbonaceous chondrites that contain all of these constituents. If not overprinted or remagnetized, chondrules and refractory inclusions are expected to carry pre-accretional magnetic records, while the matrix would recorded a post-accretional magnetic record obtained during parent-body alteration. Paleomagnetic measurements conducted with bulk samples are likely to provide magnetic records from the matrix material, because if chondrules and refractory inclusions are randomly oriented in the bulk and, even if magnetized, the magnetic moments of the inclusions cancel out and the matrix would dominate the magnetic record (see appendix). 

We report on paleomagnetic measurements of bulk matrix-rich material of two carbonaceous chondrites from the CO group. A previous study of the CV chondrite Kaba provided an upper limit on the magnetic field at 3-7 AU of <0.3 µT at $4.2_{-0.7}^{+0.8}$ Myr after CAI formation \citep{doyle_early_2014, gattacceca_new_2016}. We further investigate this region by targeting another group of carbonaceous chondrites that likely acquired a post-accretional magnetic record in this region but that, unlike Kaba, are much less likely to carry thermally induced parent-body magnetic overprints.  Assuming that magnetite could have acquired its magnetic record during alteration in the parent-body, we show that the matrices of the CO chondrites carry no stable natural remanent magnetization (NRM) despite having high fidelity magnetic recording properties and ferromagnetic minerals that formed after accretion. We show that the lack of magnetization in the matrix of the CO chondrites, together with previous paleomagnetic measurements, suggests that the dissipation of the outer solar system nebula occurred by $\sim$4.2 Myr after CAI formation. By comparing the inner and outer solar system lifetimes of the nebula, we find that the outer solar nebula dissipated <1.5 Myr after the dissipation of the inner solar nebula, suggesting that our solar nebula passed through a dual-timescale evolution. We also discuss future work necessary to obtain robust records of the solar nebula by conducting paleomagnetic and rock magnetic measurements of bulk chondrites. This includes obtaining more high-precision ages from relevant meteorites and conducting experimental work to establish the mechanisms for acquisition of magnetic field records during parent-body alteration, including in particular if secondary magnetite can record CRMs imparted by the solar nebula field.

\section{Methodology}
\subsection{Samples}
In this study, we focused on the type 3.03 CO chondrite Allan Hills (ALH) A77307 \citep{bonal_organic_2007} and on the type 3.00 CO chondrite Dominion Range (DOM) 08006 \citep{davidson_mineralogy_2019}. We selected these samples for several reasons. First, they formed magnetite during aqueous alteration on their parent body, so can potentially provide constraints on the intensity of the local ambient magnetic field present during this process \citep{doyle_early_2014}. Second, they have ferromagnetic inclusions (e.g., magnetite; see below) with high-fidelity paleomagnetic properties formed during parent-body alteration \citep{alexander_mutli-technique_2018, davidson_mineralogy_2019}. Third, they experienced very minor heating after parent-body aqueous alteration, thereby better preserving a magnetic record dating to the time of this early alteration rather than slightly later during subsequent metamorphic heating and cooling \citep{davidson_mineralogy_2019}. 

\subsubsection{CO Carbonaceous Chondrites}
Subsamples ALHA77307,157 (0.57 g) and DOM 08006,102 (2.8 g) were obtained from the US Antarctic Meteorite program (ANSMET). These were chipped from the meteorite at NASA Johnson Space Center (JSC) and did not contain any faces from band sawing at JSC that could have resulted in thermal remagnetization during cutting [e.g., \cite{mighani_end_2020, tikoo_two-billion-year_2017}]. Each contained fusion crust on one of their faces. 

Both meteorites experienced aqueous alteration and metamorphism on their parent body as indicated by the presence of only a small fraction of metal (~2 vol.\%) in the matrix and on the edges of chondrules, with most of the metal having been altered to magnetite \citep{bonal_thermal_2016, davidson_mineralogy_2019}. ALHA77307 experienced minor terrestrial weathering (Ae grade) and is estimated to have experienced a peak parent body metamorphic temperature between 200-300 $^o$C \citep{bonal_thermal_2016} resulting from aqueous alteration and metamorphism on the parent-body. DOM 08006 has experienced minor terrestrial weathering (A/B grade) \citep{davidson_mineralogy_2019} and has experienced similar or slightly lower peak temperatures in the parent body than ALHA77307 \citep{alexander_mutli-technique_2018}. ALHA77307 and DOM 08006 did not exceed peak shock pressures of 5 GPa \citep{scott_shock_1992, li_formation_2021}. Overall, these observations suggest that ALHA77307 and DOM 08006 potentially carry records of magnetic fields from the early solar system acquired during parent-body aqueous alteration. 

The paleomagnetism of the CO chondrites has been briefly studied before [see \citep{weiss_paleomagnetic_2010} for complete list of magnetic measurements with CO chondrites]. A previous study conducted thermal demagnetization of saturation magnetization of four CO3 chondrites \citep{herndon_thermomagnetic_1976}, which found that one contains magnetite as the major magnetic phase, while the remaining samples contained magnetite along with iron metal as the main magnetic phases. Two previous exploratory paleomagnetic studies obtained paleointensities from CO chondrites. Measurements of bulk subsamples of the CO3 chondrite Acfer 333 identified a low coercivity component (0-15 mT) associated with an isothermal remanent magnetization (IRM) (likely from a collectors’ hand magnet) and a high coercivity component (30-120 mT) with a paleointensity of 6 µT \citep{gattacceca_toward_2004}. Another study of bulk subsamples of the CO3.05 Yamato 81020 meteorite reported a high blocking temperature (320-640 $^o$C) component with a paleointensity of 9 $\mu$T \citep{nagata_magnetic_1991}. A paleomagnetic study of dusty olivine chondrules from ALHA77307 and DOM 08006 suggest that these meteorites have not been remagnetized since accretion to the parent-body due to their random high coercivity magnetic directions among mutually-oriented chondrules \citep{borlina_paleomagnetic_2021}. Here we focus on using ALHA77307 and DOM 08006 to conduct a detailed paleomagnetic study of bulk samples.

Previous mineralogical and petrological studies of the matrix of DOM 08006  have reported the presence of the ferromagnetic minerals kamacite (Fe$_x$Ni$_{1-x}$; with $x$ = 92-93), magnetite and pyrrhotite (Fe$_{1-x}$S; with $x$ $\sim$ 0) \citep{alexander_mutli-technique_2018, davidson_mineralogy_2019, schrader_fes_2021}. Similar minerals have been reported for ALHA77307 \citep{alexander_mutli-technique_2018, davidson_mineralogy_2019, grossman_onset_2005, schrader_fes_2021}. The amount of magnetite (6-8 wt.\%) is higher than sulfide (2-3 wt.\%) and metals (1-2 wt.\%) in the matrix of DOM 08006 \citep{alexander_mutli-technique_2018}. Below we discuss why magnetite is likely to dominate the magnetic record of these samples. Recent FORC analysis of DOM 08006 shows a tri-lobate geometry which is consistent with the magnetic carriers being predominately in vortex states \citep{sridhar_constraints_2021}. Because magnetite is the most abundant magnetic recorder and with the highest susceptibility (see below), we conclude that the magnetite grains are dominantly in the vortex state, such that they are robust magnetic recorders with relaxation times greater than the age of the solar system \citep{nagy_thermomagnetic_2019}

Because magnetite formed during parent body alteration \citep{davidson_mineralogy_2019} and was not subsequently heated above 300 $^o$C, the magnetic record of these samples should be predominately in the form of crystallization remanent magnetization (CRM). No radiogenic ages have been reported for ALHA77307 and DOM 08006. Previous work determined a $^53$Mn-$^53$Cr age from aqueously formed fayalite of 5.1$_{-0.4}^{+0.5}$ (2$\sigma$) Myr after CAI-formation from the meteorite MacAlpine Hills (MAC) 88107, a chondrite that formed from precursor material that had isotopic similarities to that of the CO parent-body but has not previously been officially categorized as a CO chondrite \citep{doyle_early_2014, torrano_relationship_2021}. As such, this measured age is likely a proxy for the youngest possible age of magnetite formation on the CO chondrite parent body. Furthermore, Al-Mg ages of chondrules from CO chondrites suggest that they formed at 2.1 $\pm$ 0.8 Myr after CAI formation \citep{borlina_paleomagnetic_2021, kita_evolution_2011} and recent modelling to explain the inventory of CAIs among different chondrite groups suggests that the CO chondrite parent-body accreted at ~2.7 Myr after CAI formation \citep{desch_effect_2018}. Thus, we can place a lower limit on the alteration age of the CO chondrites of 2.7 Myr after CAI formation, while an upper limit can be obtained by using MAC 88107’s alteration age of 5.1 Myr after CAI formation. 

\subsubsection{Formation Region of the CO Carbonaceous Chondrites}
The existence of a dichotomy in various stable isotopic systems \citep{kruijer_great_2020, scott_isotopic_2018} provides evidence that carbonaceous and non-carbonaceous meteorites formed in two distinct reservoirs. Although the exact locations for these reservoirs are unclear, it is generally thought that carbonaceous chondrites likely formed at >3 AU \citep{brasser_partitioning_2020, demeo_solar_2014, desch_effect_2018, kruijer_great_2020, morbidelli_dynamical_2015}. Measurements of water contents from CI, CM, CR and LL chondrites provide evidence that carbonaceous chondrites formed at <7 AU \citep{sutton_bulk_2017}. Thus, we assume the formation region of the carbonaceous chondrites to be between 3-7 AU. We also assume that chondrite parent bodies originate from the midplane region of the PPD because drag forces on 0.1-mm-sized grains lead them to settle on the midplane of the PPD, resulting in the formation of the parent-bodies in that region \citep{weiss_history_2021}. 

\subsubsection{Magnetite as the Main Magnetic Carrier of Bulk Samples}
Because bulk samples contain magnetic minerals that are pre-accretional (e.g., kamacite) and post-accretional (e.g., magnetite) in origin, it is important to establish which one of them will dominate the magnetization record. The NRM moment $M$, measured in the laboratory, is proportional to the magnetic field, $B$ experienced by the sample, through:
\begin{equation}
    M = \frac{\chi_R}{\mu_{0}} B, 
\end{equation}

where $\chi_{R}$ is the remanent susceptibility of the sample \citep{dunlop_rock_1997} and $\mu_0$ is the vacuum permeability. We can write $\chi_{R}$ as \citep{kletetschka_fundamental_2017}:

\begin{equation}
    \chi_R \sim \frac{M_{rs}}{c M_s} = \frac{s}{c},
\end{equation}

where $M_{rs}$ s the saturation remanent magnetization, $M_s$ is the saturation magnetization, $s = M_{rs}/M_s$ is the squareness ratio, and $c$ is a s a grain-shape dependent constant. Thus, for a sample that contains magnetite and kamacite, we can determine $\chi_R$ based on the sum of the remanent susceptibilities of each mineral through:

\begin{equation}
    \chi = \chi_{mag} + \chi_{kam},
\end{equation}

\begin{equation} \label{chimag}
    \chi_{mag} \sim s_{mag},
\end{equation}

\begin{equation}\label{chikam}
    \chi_{kam} \sim s_{kam},
\end{equation}

where $\chi_{mag}$ and $s_{mag}$ are the remanent susceptibility and squareness of the magnetite and $\chi_{kam}$ and $s_{kam}$ are the remanent susceptibility and squareness of the kamacite, where we have assumed the grain-shape dependent constants for magnetite and kamacite are similar ($c_{kam}\sim c_{mag}$). Taking the ratio between Eqs. \ref{chimag} and \ref{chikam}, we have:

\begin{equation}
    \frac{\chi_{mag}}{\chi_{kam}} = \frac{s_{mag}}{s_{kam}},
\end{equation}

Because most magnetite is likely to be in the single-vortex and single-domain state while kamacite is likely to be multi-domain, we can assume $s_{mag}/s_{kam}>100$ \citep{dunlop_theory_2002}. Thus, $\chi_{mag} > 100 \chi_{kam}$, indicating that magnetite will be the main magnetic carrier of the NRM in these samples.

\subsection{Paleomagnetic and Rock Magnetic Experiments}
Sample preparation and paleomagnetic analyses were conducted in the Massachusetts Institute of Technology (MIT) Paleomagnetism Laboratory. Samples from each meteorite were obtained by cutting with a wire-saw cooled with ethanol to avoid acquisition of partial TRM (pTRM) during cutting. We obtained eight mutually oriented samples from ALHA77307 and eleven mutually oriented samples from DOM 08006. Tables \ref{bulkdirec1} and \ref{bulkdirec2} summarizes the mass and distance from the fusion crust to the outer edge of each sample. Paleomagnetic measurements were obtained in a magnetically shielded room (residual field $\sim$200 nT) using a 2G Enterprises Superconducting Rock Magnetometer (SRM) 755R equipped with an automatic alternating field (AF) three-axis degausser and remagnetization system \citep{kirschvink_rapid_2008}. The three-axis moment noise level of the MIT 2G SRM is <1×10$^{−12}$ Am2 \citep{wang_lifetime_2017}.

We AF demagnetized four samples from ALHA77307 (one containing fusion crust and the remaining from >1.8 mm into the interior) and eight samples from DOM 08006 (one containing fusion crust and the remaining from >1 mm into the interior). These samples were glued with cyanoacrylate cement onto non-magnetic quartz holder with magnetic moments of <4$\times$10$^-12$ Am$^2$. Samples were demagnetized with AF up to 145 mT to identify NRM components. We measured the magnetic moment after the AF application in each one of three directions and averaged the measurements to reduce spurious anhysteretic remanent magnetization (ARM) and gyroremanent magnetization (GRM) \citep{stephenson_three-axis_1993}. We also applied ARMs (AC field of 260 mT and DC bias field of 50 µT) to these samples and AF demagnetized them up to 145 mT.  Paleointensities were estimated for these samples using the ARM method \citep{tikoo_decline_2014}. 

We thermally demagnetized four samples of ALHA77307 (two containing fusion crust and two interior samples) to identify NRM components and obtain additional paleointensity estimates. We followed the in-field zero-field zero-field in-field (IZZI) double heating protocol to obtain paleointensities \citep{yu_toward_2004}. Heating measurements were performed using an ASC Scientific TD48-SC thermal demagnetizer in air with thermal accuracy better than $\pm$5 $^o$C. Samples were heated to the set temperature for 20 minutes and cooled within 40 minutes to room temperature. For ALHA77307 and DOM 08006, we initiated the IZZI protocol at 325 $^o$C and 50 $^o$C with steps below this being only thermally demagnetized (i.e., only heating in zero-field). The IZZI protocol was conducted in steps of 50 $^o$C, with a 50 $\mu$T bias field and partial TRM (pTRM) checks in 100 $^o$C steps to estimate whether thermochemical alteration could have occurred during the measurements. Samples were taped to non-magnetic quartz holders for measuring their magnetic moment. Initially, three interior samples of DOM 08006 were thermally demagnetized. However, two of these broke into two mutually oriented subsamples each at 350 $^o$C, leading to five thermally demagnetized samples. We conducted the IZZI protocol with two of these five samples. For three samples of ALHA (two fusion crusted and one interior) that we conducted IZZI experiments with, we removed low coercivity components (up to 21 mT) with AF demagnetization prior to the start of the IZZI protocol. We AF demagnetized to the same levels after each in-field step during the protocol. We also thermally demagnetized an ARM application (AC field of 145 mT and DC bias field of 50 $\mu$T) of two samples, following AF demagnetization of the NRM, to identify the Curie temperature of the ferromagnetic carriers.  

Fits for both the ARM and IZZI paleointensities were obtained either using ordinary least square (OLS) or reduced major axis (RMA) \citep{borlina_paleomagnetic_2021, smith_use_2009}. The quality of the results from the IZZI protocol were quantified using the PICRIT03 criteria \citep{paterson_improving_2014}, which quantify thermochemical alteration during the protocol. The ARM method has the advantage over the IZZI protocol of not involving laboratory heating that otherwise can lead to thermochemical alteration. Principal component analysis (PCA) was used to determine the direction and maximum angular deviation (MAD) of the magnetization components identified during the AF and thermal demagnetization \citep{kirschvink_least-squares_1980}. To identify any origin trending components, we calculated the deviation angle (DANG). For samples that DANG < unanchored MAD, we calculated the PCA with an anchored fit \citep{tauxe_strength_2004}. 

Nonetheless, because the non-fusion-crusted samples contain CRMs, both the ARM and the IZZI methods cannot reliably reproduce in the laboratory the natural process by which the NRM was acquired by these samples. Therefore, knowledge of the ratio of CRM to ARM and CRM to thermoremance (TRM) is required to obtain paleointensity estimates from these experiments.  Unfortunately, these have been poorly constrained by previous studies. Previous studies determined an TRM to ARM ratio $f'$ of 3.33 for magnetite and Fe-Ni bearing samples, which has an estimated 2$\sigma$ uncertainty that could be of at least a factor of 5 \citep{weiss_lunar_2014}. For alteration timescales of a year to millions of years, such as those experienced by carbonaceous chondrites \citep{dyl_early_2012, ganino_fumarolic-like_2020, krot_timescales_2006}, TRM/CRM can be $\sim$1-2 \citep{mcclelland_theory_1996}. For shorter alteration timescales ($\sim$10$^3$ s), this ratio can be as large as $\sim$5-10 \citep{mcclelland_theory_1996}. For this study, we assume that TRM$\sim$CRM and we use $f'$ = 3.33 for the ARM method. We note that the paleointensities and recording limits described here are likely to be an underestimations and future experiments are necessary to fully determine the CRM to ARM ratio. 

We sought to estimate the minimum paleointensity that we could recover from our samples using the ARM method \citep{bryson_paleomagnetic_2017, tikoo_magnetic_2012}. We determined this for each sample by applying a sequence of ARMs (AC field of 145 mT) with DC bias fields ranging from 1 to 10 $\mu$T.  We then attempted to retrieve these paleointensities using the ARM method described above (using an ARM bias field of 50 $\mu$T) in the same coercivity range used to determine the paleointensity. We quantified the accuracy and precision of the retrieved paleointensity estimates with deviation of the retrieved magnetic field from the applied field ($D'$) and the uncertainty of the retrieved field ($E$) metrics \citep{bryson_paleomagnetic_2017, tikoo_magnetic_2012}. Again, we assumed $f'$ = 3.33 such that the range of applied CRM-equivalent paleointensities were 0.3 to 3 $\mu$T. For a given sample, we estimate the minimum paleointensity that can be recovered as the lowest value for which a sample having -0.5<$D'$<1 and $E$<1. 
We also experimentally determined the viscous remanent magnetization (VRM) acquisition and decay rate of one of our samples from DOM 08006 to determine what fraction of the NRM could have been acquired through VRM. We exposed a previously demagnetized sample for 37 days in a terrestrial field of 46 $\mu$T and measurements were taken periodically during this time. VRM decay was measured while the sample was kept inside of a magnetically shielded room ($\sim$200 nT) for 14 days and measured periodically.  Linear fits were used to calculate the VRM acquisition and decay rates.  

\section{Results}

\subsection{ALHA77307}
\begin{figure}[t!]
\centering
  \includegraphics[scale=0.5]{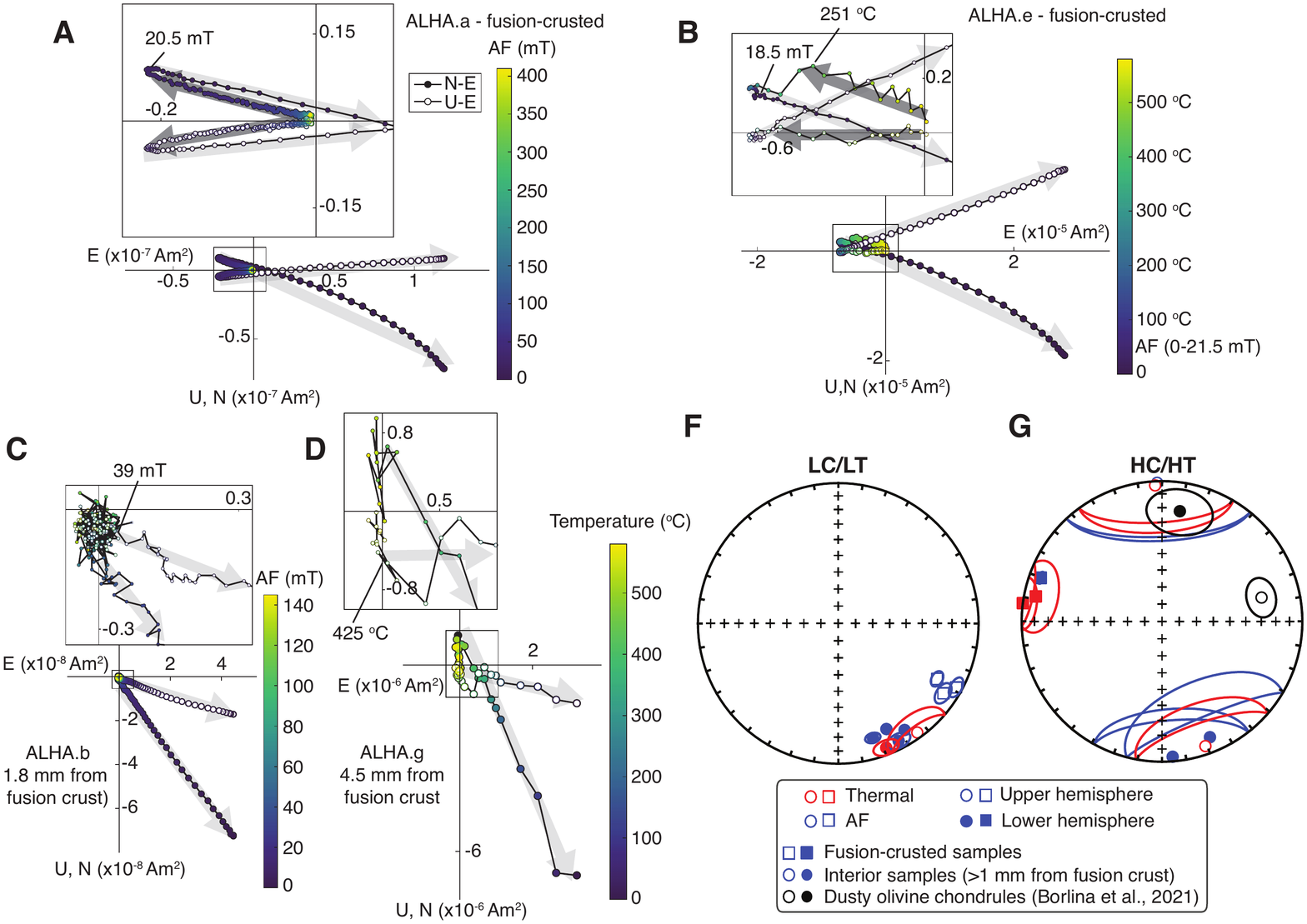}
  \caption{\textbf{AF and thermal demagnetization of samples from ALHA77307.} Selected orthographic projections of NRM vector endpoints during AF and thermal demagnetization for samples (A) ALHA.a, (B) ALHA.e, (C) ALHA.b and (D) ALHA.g. Closed symbols show the north-east (N-E) projection of the magnetization and open symbols show the up-east (U-E) projection of the magnetization. Selected AF and thermal steps are labeled. (A) and (C) show AF demagnetization from fusion-crusted and interior samples (>1 mm away from fusion crust), respectively. (B) and (C) show thermal demagnetization from fusion-crusted and interior samples, respectively. (F-G) show equal area stereonets with the directions of low coercivity/low temperature (LC/LT) and high coercivity/high temperature (HC/HT) components from all samples measured from ALHA77307. Open and closed symbols represent upper and lower hemispheres, respectively. Red symbols show components from thermal demagnetization, while blue symbols show components from AF demagnetization. Squares show fusion-crusted samples, while circles show interior samples. Black datapoints in equal area stereonet show data from HC components from dusty olivine chondrules from \cite{borlina_paleomagnetic_2021}.}
  \label{fig:bulkALHA}
\end{figure}

All AF-demagnetized samples from both the interior and with fusion crust contained a low coercivity (LC) component (blocked up to 58 mT) that had directions with overlapping MADs (Fig. \ref{fig:bulkALHA}; Table \ref{bulkdirec1}). The fusion-crusted samples also contained a high coercivity (HCf) component blocked between 26-410 mT (Fig. \ref{fig:bulkALHA}). By comparison, further AF demagnetization of the interior samples to 145 mT demonstrated that they contained no stable magnetization after removal of the LC component (Fig. Figs. \ref{fig:bulkALHA} and see supplemental material). Using the ARM method, the fusion-crusted samples yielded an apparent paleointensity of 358.5 $\pm$ 15.8 $\mu$T (these and all other reported uncertainties are 2$\sigma$) for the LC component and a paleointensity of 26.7 $\pm$ 0.8 $mu$T for the HCf component (see supplemental material). The presence of a common LC component among all four samples and the high apparent paleointensity retrieved from the LC component suggest that ALHA77307 was exposed to contamination from a magnet (i.e., IRM) after atmospheric entry. We note that the fact that the fusion crusted samples were not completely remagnetized (i.e., they have both LC IRM overprint and the HCf component) and that the paleointensity of the HCf component has a strength similar to Earth’s magnetic field implies that the HCf component is likely a TRM acquired during atmospheric entry. The presence of a HCf component in the fusion-crusted samples and not in interior samples demonstrates that ALHA77307 was not completely remagnetized by the IRM.

\begin{figure}[t!]
\centering
  \includegraphics[scale=0.5]{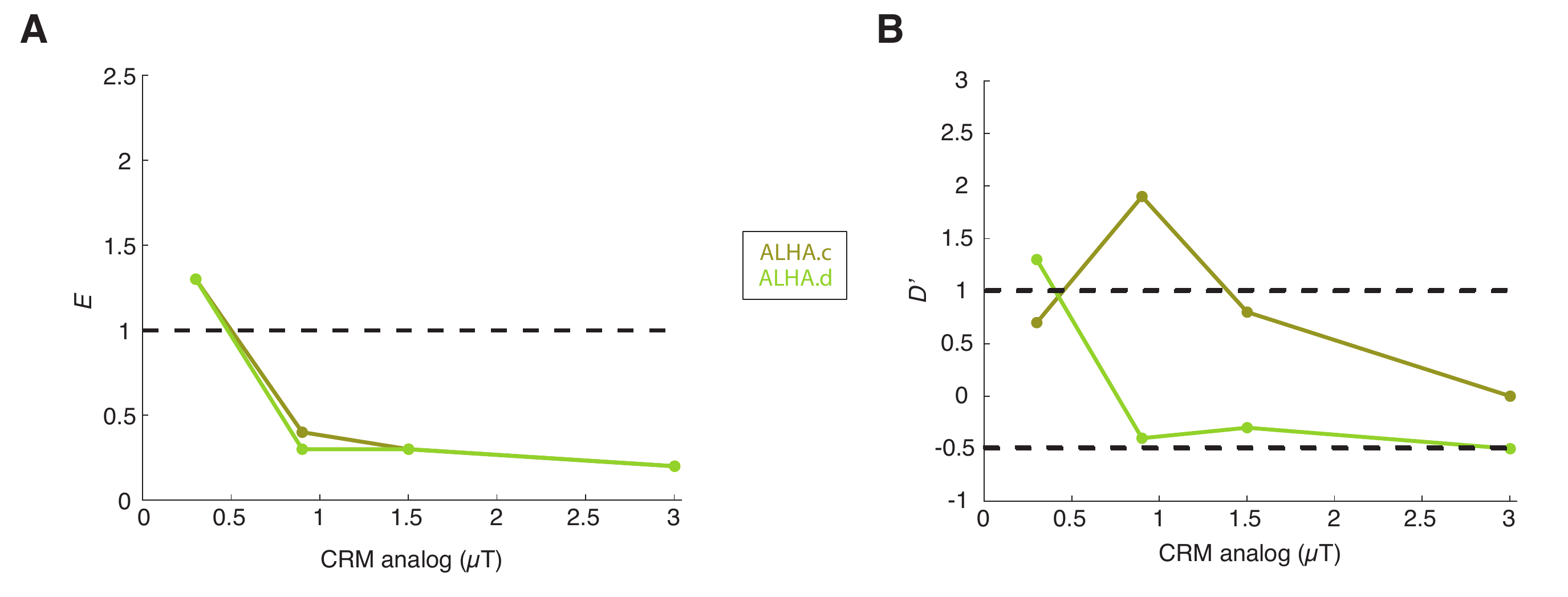}
  \caption{\textbf{$E$ and $D'$ values calculated for samples from ALHA77307.} Values below (A) and between (B) the dashed lines represent reliable paleomagnetic fidelities. The coercivity range, of each sample, used to calculate the paleointensities were used to calculate the E and D’ values. The CRM analogs were calculated using $f'$ = 3.33 from ARMs (AC field of 145 mT) with DC bias fields of 10, 5, 3 and 1 $\mu$T (estimated CRM-equivalent fields of 3, 1.5, 0.9 and 0.3 $\mu$T).}
  \label{fig:bulkfidelityALHA}
\end{figure}

The ARM method applied to the apparently non-magnetized high coercivity (HC) range of the interior samples (see supplemental material) yielded a paleointensity of 2.2 $\pm$ 0.9 $\mu$T (Table \ref{bulkARM}). Although the 2$\sigma$ uncertainty does not formally encompass zero paleointensity, the low paleointensity value and the lack of a unidirectional and stable component in the HC range (see supplemental material) suggest that a weak or null field was present when ALHA77307 acquired its magnetic record. Fig. \ref{fig:bulkfidelityALHA} and Table \ref{bulkARM} show that the interior sample (ALHA.d) can record fields >0.9 $\mu$T.

Thermal demagnetization of a 50 $\mu$T ARM applied to two of the interior samples that were previously AF demagnetized, show that the magnetic remanence is carried by sulfide, magnetite (the main magnetic carrier; see above), and Fe-metal, likely awaruite (see supplementary material), consistent with the expected mineralogy of CO chondrites. The results of the IZZI protocol are presented in Table \ref{bulkTHERMAL} and in the supplemental material. After the removal of the LC component from the fusion crusted samples, we observed high temperature (HTf) components (250-580 $^o$C) oriented close to the directions of the HCf components (Fig. \ref{fig:bulkALHA}). The fusion-crusted samples successfully passed the PICRIT03 criteria, indicating minimum thermochemical alteration during laboratory heating. The paleointensities retrieved from the HTf components were of 13.9 $\pm$ 4.7 $\mu$T and 13.9 $\pm$ 5.4 $\mu$T (Table \ref{bulkTHERMAL}; see supplemental material). These are within error of the ARM paleointensities retrieved from the HCf component given the uncertainties in the values of ARM/CRM and TRM/CRM. One interior sample (ALHA.g) had a LT component (0-425 $^o$C) similar in direction to the LC components of the other samples followed by an origin-trending HT component (425-580 $^o$C) oriented close to the direction of the laboratory field. The latter may have resulted from thermochemical alteration during the heating experiments (Fig. \ref{fig:bulkALHA}D). The other interior sample (ALHA.h; LC removed with AF prior to the IZZI protocol) also had a LT component (51-475 $^o$C) oriented in the direction of the LC components of the other samples (see supplemental material). Due to thermochemical alteration during heating experiments, both interior samples fail multiple PICRIT03 alteration selection criteria [Table \ref{bulkTHERMAL}; \cite{paterson_improving_2014}], which implies that the HT component paleointensities of the interior samples cannot be reliably estimated. 
In summary, the LC/LT component common to all samples is likely an IRM that did not completely overprint the sample, almost certainly acquired during sample handling. The HCf/HTf component, which is only present in the fusion-crusted samples, is consistent with magnetization acquired during heating from atmospheric entry in Earth’s magnetic field. The lack of a unidirectional HC magnetization among most interior samples together with fidelity tests indicate that the ancient field environment was less <0.9 µT when ALHA77307 acquired its magnetization.

\subsection{DOM 08006}
All AF-demagnetized DOM 08006 samples from the interior had a common LC component blocked between 6-28 mT (Figs. \ref{fig:bulkDOM} and see supplemental material; Table \ref{bulkdirec2}). The fusion-crusted sample had a single origin-trending HCf component blocked up to 145 mT (Fig. \ref{fig:bulkDOM}). Using the ARM method, we obtained a paleointensity of 39.5 $\pm$ 0.4 $\mu$T for the HTf component of the fusion-crusted sample, consistent with Earth’s magnetic field strength as recorded during atmospheric entry (Table \ref{bulkARM}; Fig. \ref{fig:bulkDOMARM}). 

\begin{figure}[t!]
\centering
  \includegraphics[scale=0.5]{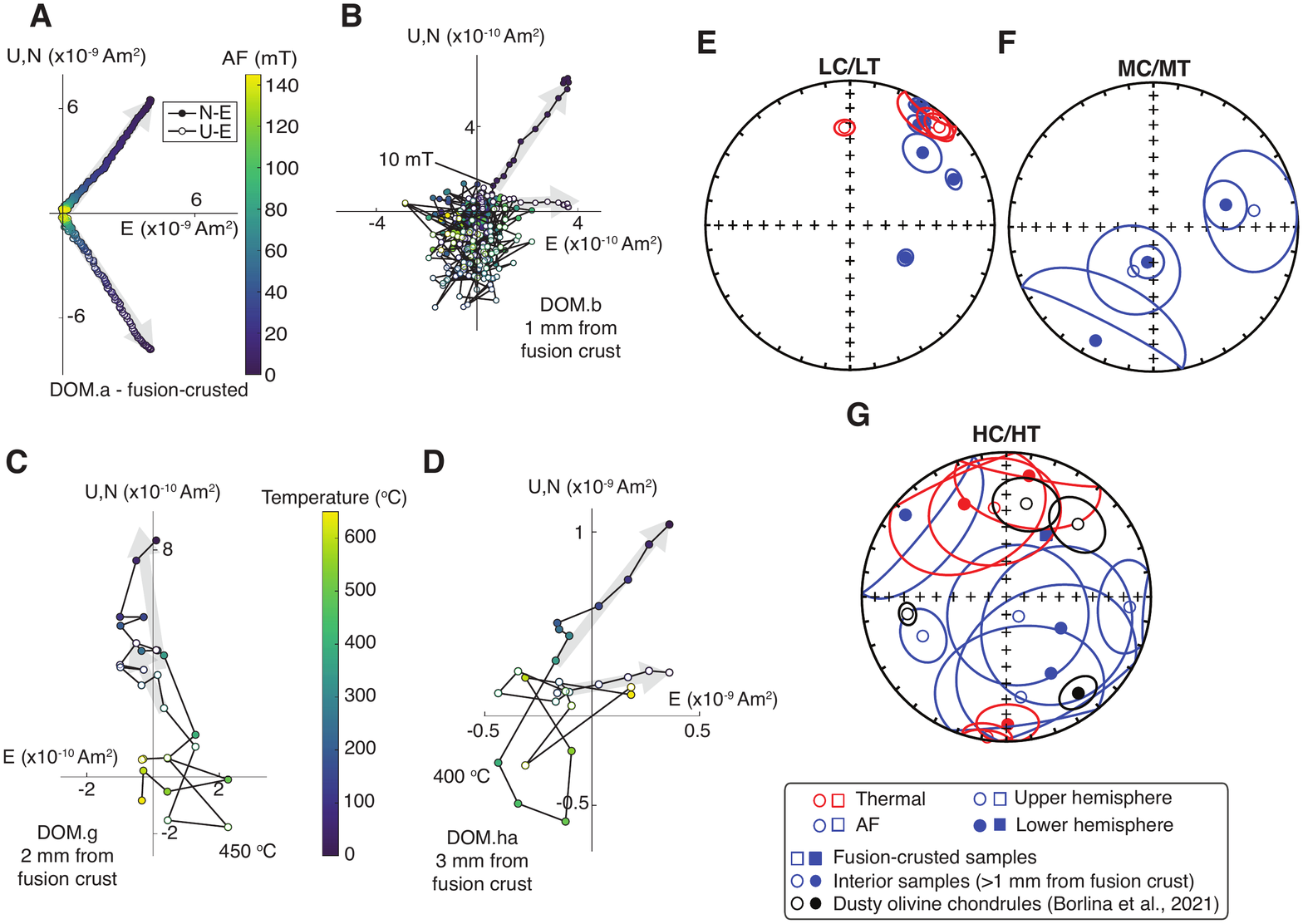}
  \caption{\textbf{AF and thermal demagnetization of samples from DOM 08006.} Selected orthographic projections of NRM vector endpoints during AF and thermal demagnetization for samples (A) DOM.a, (B) DOM.b, (C) DOM.g and (D) DOM.ha. Closed symbols show the north-east (N-E) projection of the magnetization and open symbols show the up-east (U-E) projection of the magnetization. Selected AF and thermal steps are labeled. (A-B) show AF demagnetization from fusion-crusted and interior samples (>1 mm away from fusion crust), respectively. (C-D) show thermal demagnetization from interior samples. (E-G) show equal area stereonets with the directions of low coercivity/low temperature (LC/LT), medium coercivity/medium temperature (MC/MT) and high coercivity/high temperature (HC/HT) components from all samples measured from DOM 08006. Open and closed symbols represent upper and lower hemispheres, respectively. Red symbols show components from thermal demagnetization, while blue symbols show components from AF demagnetization. Squares show fusion-crusted samples, while circles show interior samples. Black datapoints in equal area stereonet show data from HC components from dusty olivine chondrules from \cite{borlina_paleomagnetic_2021}.}
  \label{fig:bulkDOM}
\end{figure}

Four interior samples had medium coercivity (MC) components with coercivities up to 45 mT (Fig. \ref{fig:bulkDOM}), but no stable HC magnetization blocked beyond this. A fifth sample had a HC component between 19-145 mT (see supplemental material). The remaining two samples did not carry any components blocked above 15 mT.  PCA of the HC range yields scattered directions and MAD values >40$^o$ for most samples (Fig. \ref{fig:bulkDOM}; Table \ref{bulkdirec2}). The LC and MC components are likely associated with viscous remanent magnetization (VRM) and/or low temperature parent body thermochemical alteration. To assess this, we analyzed the VRM acquisition and decay properties of sample DOM.m following the AF demagnetization experiments. We measured a VRM acquisition rate of of 1.15$\times$10$^{-6}$ Am$^2$ kg$^{-1}$ $\mu$T$^{-1}$ and a VRM decay rate that was shallow at first ($\sim$10$^3$ seconds) with a decay rate of 2.06$\times$10$^{-8}$ Am$^2$ kg$^{-1}$ $\mu$T$^{-1}$, followed by an increase in the decay rate to a value of 8.17$\times$10$^{-7}$ Am$^2$ kg$^{-1}$ $\mu$T$^{-1}$ (Fig. \ref{fig:bulkDOMVRM}). Because Antarctic meteorites typically have terrestrial ages of a few tens of thousands of years (Jull, 2006) with some surviving for up to 2 million years, we estimate that $\sim$52\% of the NRM of the DOM.m could have been acquired in 10,000 years on the Earth’s surface, consistent with the magnetic moment carried by the LC component. 

The ARM method yielded a mean paleointensity of 1.2 $\pm$ 1.4 $\mu$T for the HC range (Figs. 3 and S6; Table 2). Like the interior samples from ALHA77307, the lack of stable, unidirectional components in the HC range, the paleointensities within error of zero retrieved from the ARM method, and the demonstrated capacity for gaining ARM in the same non-magnetized range (Figs. \ref{fig:bulkDOMARM} and see suplemenetal material) indicate that a weak to null field was present when the magnetic record was acquired. We chose to not include DOM.c and DOM.l as part of our paleointensity analysis. DOM.c contains a well constrained HC component (19-145 mT, MAD = 13.9$^o$) that is not observed in any other sample and records a paleointensity of 4.1 $\pm$ 0.2 $\mu$T. DOM.l also contained a HC component (28-145 mT, MAD = 21$^o$) with a distinct direction than that of DOM.c, which yield a paleointensity of 3.4 $\pm$ 0.4 $\mu$T, which is higher than its fidelity limit of 0.9 $\mu$T (Fig. \ref{fig:bulkfidelityDOM}; Table \ref{bulkARM}). These records are not observed in any of the other subsamples and may be due to the presence of large or highly magnetic chondrules and refractory inclusions in the bulk samples with pre-accretional magnetization and/or with spurious remanence acquired during AF demagnetization \citep{weiss_paleomagnetic_2010}. Not including DOM.c and DOM.l results in a mean paleointensity of 0.2 $\pm$ 0.4 $\mu$T. The fidelity test for interior samples from DOM 08006 (Fig. \ref{fig:bulkfidelityDOM}; Table \ref{bulkARM}) show that the sample with best fidelity record is DOM.j, recording fields >0.9 µT. We conclude that the HC magnetic record was likely acquired in an environment with paleointensity of <0.9 µT. 

The thermal experiments identified LT components (blocked up to 350 $^o$C) in all samples whose direction is like those of the direction of the LC components (Fig. \ref{fig:bulkDOM}). No components were observed blocked above 350 $^o$C as indicated by a scattered directions and large MADs yielded by PCA fits (Fig. \ref{fig:bulkDOM}; Table \ref{bulkdirec2}). The two samples on which we conducted the IZZI experiment failed multiple PICRIT03 alteration selection criteria. Thus, paleointensities and components direction are unlikely to be robust (see supplemental material; Table \ref{bulkTHERMAL}). 

In summary, the LC/LT/MC/MT components can be associated with low temperature parent body thermochemical alteration and/or VRM acquisition and the fidelity test indicates that the HC range of DOM 08006 recorded a field <0.9 µT. 

\begin{figure}[t!]
\centering
  \includegraphics[scale=0.5]{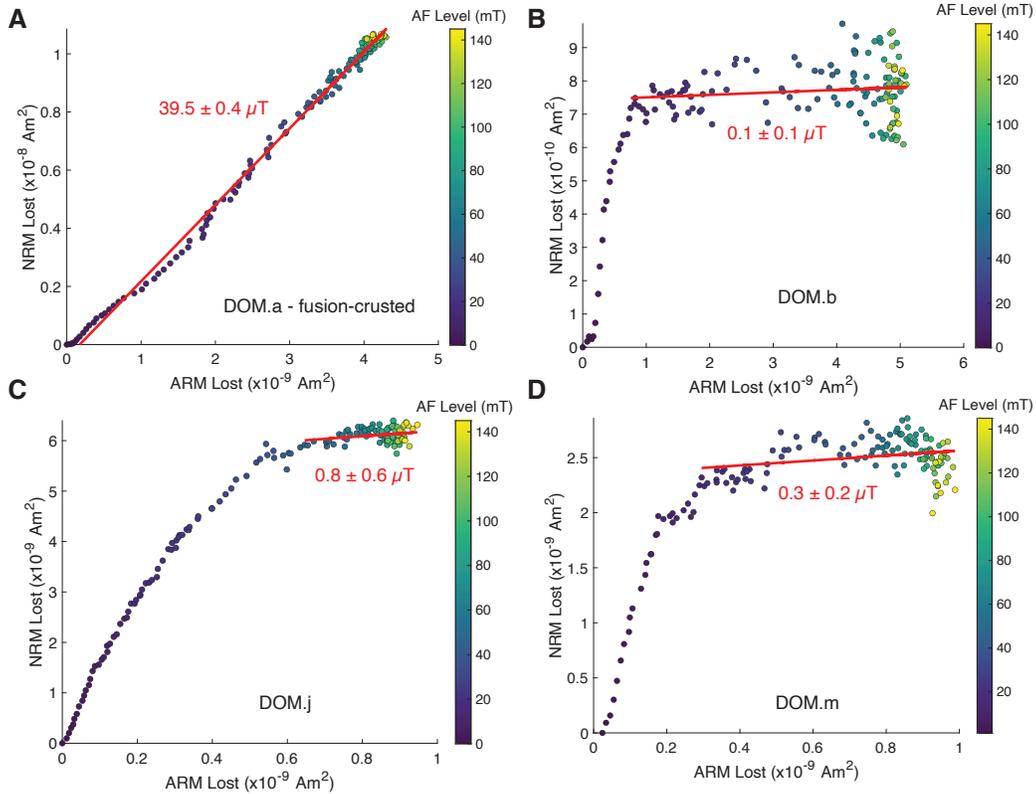}
  \caption{\textbf{ARM paleointensity experiments for the selected samples (A) DOM.a, (B) DOM.b, (C) DOM.j and (D) DOM.m from DOM 08006.} Shown is NRM lost during stepwise AF demagnetization up to 145 mT and ARM lost during AF demagnetization up to 145 mT of an ARM (AC field of 145 mT with a DC bias field of 50 µT). Paleointensitites ($f'$ = 3.33) and their 95\% confidence intervals are reported. The red lines represent the range of coercivities used to calculate the fit.}
  \label{fig:bulkDOMARM}
\end{figure}

\begin{figure}[t!]
\centering
  \includegraphics[scale=0.5]{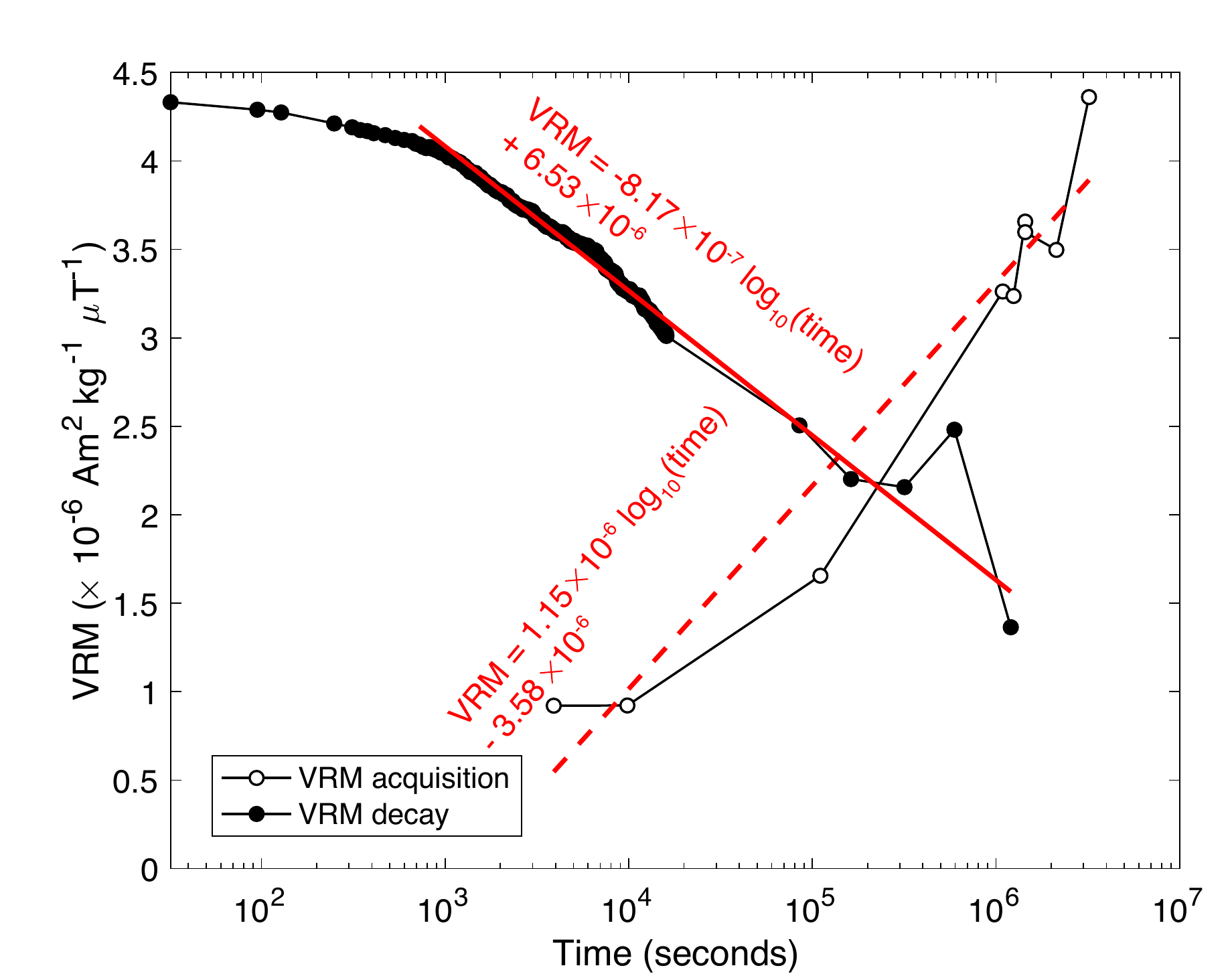}
  \caption{\textbf{VRM acquisition and decay over a period of $\sim$40 days of DOM.m, a 1.71 mg sample from DOM 08006.} Closed show measurements of the VRM decay experiment, open symbols show measurements of the VRM acquisition experiment, solid line shows linear fit of the VRM decay experiment, and dashed line shows linear fit of the VRM acquisition experiment.}
  \label{fig:bulkDOMVRM}
\end{figure}

\begin{figure}[t!]
\centering
  \includegraphics[scale=0.5]{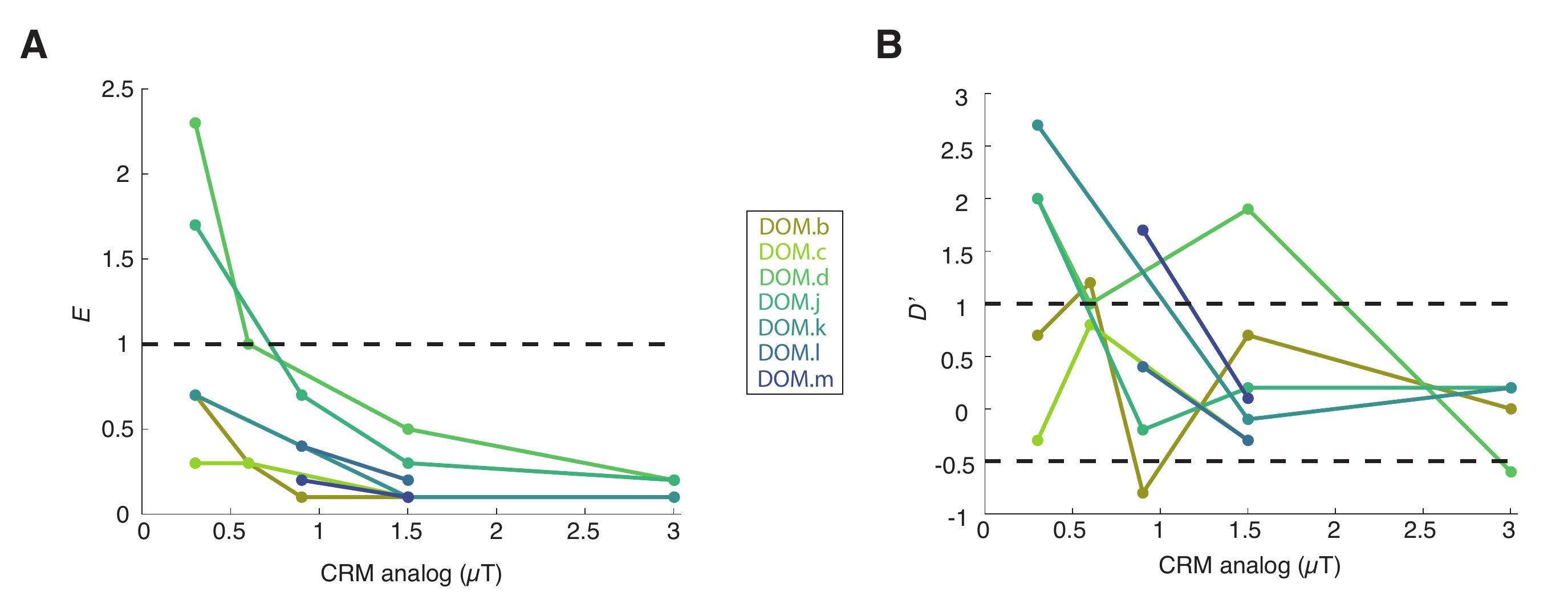}
  \caption{\textbf{$E$ and $D'$ values calculated for samples from DOM 08006.} Values below (A) and between (B) the dashed lines represent reliable paleomagnetic fidelities. The coercivity range, of each sample, used to calculate the paleointensities were used to calculate the E and D’ values. The CRM analogs were calculated using $f'$ = 3.33 from ARMs (AC field of 145 mT) with DC bias fields of 10, 5, 3 and 1 $\mu$T (estimated CRM-equivalent fields of 3, 1.5, 0.9 and 0.3 $\mu$T). }
  \label{fig:bulkfidelityDOM}
\end{figure}

\section{Discussion}
In the early solar system, there were at least four sources of magnetic fields that could conceivably have exceeded 0.9 µT: the solar wind magnetic field; impact-generated fields; a parent-body core dynamo magnetic field; and the solar nebula magnetic field. Below we discuss the implications of the magnetic record obtained from the CO chondrites and previous measurements for each one of these possibilities. We note that this discussion assumes that the magnetite is capable of having recorded a CRM during parent-body alteration. We discuss below why this might not be the case and future directions to address this.  

\subsection{Implications for the Solar Wind Magnetic Fields }
Previous studies have discussed the possibility of the solar wind as a source of remanence recorded by the parent bodies of meteorites \citep{obrien_arrival_2020, oran_were_2018}. MHD simulations indicate that the solar wind magnetic field could instantaneously have been amplified to a few µT, but the time-averaged field that would be recorded as a CRM in our meteorites would be several orders of magnitude lower than this value \citep{oran_were_2018}). Our findings from CO chondrites place an upper limit to the time-averaged solar wind magnetic field of <0.9 µT, consistent with these studies.  

\subsection{Implications for Impact and Dynamos in Planetesimals}
Planetesimals have been proposed to have hosted magnetic fields through different mechanisms. A planetesimal dynamo could have been generated by advection of an interior liquid metallic core \citep{sterenborg_thermal_2013}. Alternatively, magnetic fields could have been produced transiently during impacts \citep{weiss_paleomagnetic_2010}. Our records could imply that at the time that the magnetic record was acquired, fields with intensities <0.9 µT (due to a dynamo, crustal remanence from a previous dynamo or impact-generated field) were present on the CO parent. Weak fields such as the upper limit of our measurements are within the values predicted by core dynamo scaling laws \citep{weiss_paleomagnetic_2010}. Furthermore, models of thermal convection-driven dynamos suggest that field generation could start by ~4.5-5 Myr after CAI-formation \citep{bryson_constraints_2019}, relatively close to the timing of magnetic acquisition of our samples. In summary, any magnetic field on the parent body at the time of CO magnetization was weaker than <0.9 µT. 

\begin{figure}[t!]
\centering
  \includegraphics[scale=0.4]{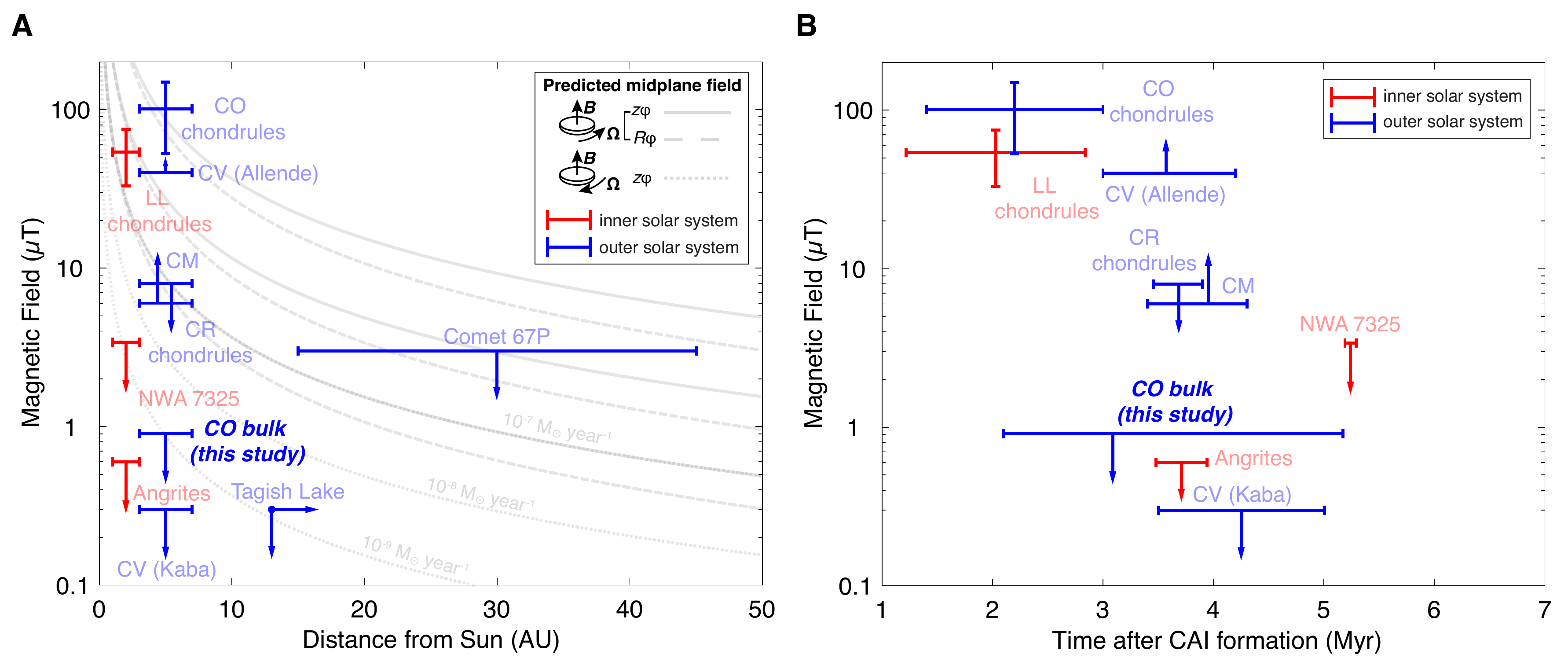}
  \caption{\textbf{Summary of previous paleomagnetic studies along with results from this study and models of the magnetic field from the solar nebula.} Points show paleomagnetic constraints from LL chondrules \citep{fu_solar_2014}, CO chondrules \citep{borlina_paleomagnetic_2021}, bulk NWA 7325 achondrites \citep{weiss_nonmagnetic_2017}, bulk angrites \citep{wang_lifetime_2017}, bulk CV (Kaba) chondrites \citep{gattacceca_new_2016}, Fe-sulfides in CV (Allende) \citep{fu_fine-scale_2021}, CR chondrules \citep{fu_weak_2020}, Rosetta observations of comet 67P Cheryumov Gerasimenko \citep{biersteker_implications_2019} and this study. Red symbols show samples that are likely from the inner solar system (<3 AU) and blue symbols show samples that are likely from the outer solar system (>3 AU). (A) Solid lines show predicted midplane magnetic field: solid lines show field due to vertical Maxwell stress [$z\varphi$; eq. 16 of \cite{bai_heat_2009}] and dashed lines show field due to radial Maxwell stress ($R\varphi$; eq. 7 of \cite{bai_heat_2009}], both assuming the nebular magnetic field and sense of disk rotation are aligned. Dotted lines show the field due to vertical Maxwell stress [$z\varphi$; eq. 7 of \cite{bai_heat_2009}] assuming nebular magnetic field and sense of disk rotation are anti-aligned (radial component cancels out for this case). All curves were calculated assuming accretion rates of 10$^{-9}$ (bottom curve), 10$^{-8}$ (middle curve) and 10$^{-7}$ (top curve) $M_\odot$ year$^{-1}$. (B) Points show paleomagnetic constraints from (A) as a function of age (see text for details on the estimates of the ages).}
  \label{fig:bulkfields}
\end{figure}

\subsection{Solar Nebular Magnetic Fields and Implications for the Dissipation of PPDs and Planetary Formation}
Previous measurements support the presence of nebular fields in the early solar system \citep{borlina_paleomagnetic_2021, cournede_early_2015, fu_fine-scale_2021, fu_solar_2014}. The youngest record of the solar nebula in the inner solar system comes from LL chondrules at 2.03 $\pm$ 0.81 Myr after CAI formation \citep{fu_solar_2014, weiss_history_2021}, and the youngest record in the outer solar system comes from the CM chondrites at 3.9$_{-0.5}^{+0.4}$ Myr after CAI formation \citep{cournede_early_2015, fujiya_evidence_2012}. We note that I-Xe ages for the CMs have been reported as 2.9 $\pm$ 0.39 Myr after CAI formation in the form of an abstract \citep{pravdivtseva_i-xe_2013} and recent $^{53}$Mn-$^{53}$Cr ages support a younger alteration time of 3.9$_{-0.5}^{+0.4}$ Myr after CAI formation \citep{fujiya_evidence_2012}. Fig. \ref{fig:bulkfields} shows the results of the CO chondrites in context with those of previous studies \citep{biersteker_implications_2019, borlina_paleomagnetic_2021, bryson_paleomagnetic_2017, bryson_evidence_2020, cournede_early_2015, fu_solar_2014, fu_weak_2020, fu_fine-scale_2021, gattacceca_new_2016, wang_lifetime_2017, weiss_nonmagnetic_2017}. We present in Fig. \ref{fig:bulkfields}A magnetic field models as a function of the distance from the Sun. Shown are three set of curves, each assuming magnetic fields drove accretion at rates ranging from 10$^-7$ to 10$^-9$ $M_\odot$ year$^{-1}$ as typically observed for actively accreting Young Stellar Objects \citep{hartmann_accretion_1998, weiss_history_2021}. The first and second curves assume radial-azimuthal and vertical-azimuthal magnetic field stresses driving accretion in the case when the nebular field and the disk angular moment are aligned, while the third curve assumes vertical-azimuthal stresses when the nebular field and the disk rotation are anti-aligned. Recent studies have supported the presence of magnetic inhomogeneities in the early solar system \citep{borlina_paleomagnetic_2021, fu_fine-scale_2021}, which suggest that magnetic fields might not decay monotonically as proposed by theoretical models of solar nebula magnetic fields.

Measurements from dusty olivine chondrules in LL ordinary chondrites \citep{fu_solar_2014}, are consistent with predictions from magnetically-driven accretion at rates between 10$^{-8}$ and 10$^{-7}$ $M_\odot$ year$^{-1}$ and provide lower limits to the accretion rate when compared to that reported by \cite{borlina_paleomagnetic_2021} from dusty olivine chondrules in CO chondrites. Comparing the accretion rate inferred from the lower limit of the paleointensity from the LL chondrules \citep{fu_solar_2014} to the models shown in Fig. \ref{fig:bulkfields}A, we obtain a value of $\sim$10$^{-8}$ $M_\odot$ year$^{-1}$. This model assumes an aligned configuration between the magnetic field and rotation of the disk. Under this configuration, a field of $\sim$10 $\mu$T would suggest that the solar nebula was still present in the outer solar system. Comparing the CO and CV bulk magnetic records with models that assume accretion rates >10$^{-8}$ $M_\odot$ year$^{-1}$, we observe that the upper limits from this study (<0.9 $\mu$T) and that of Kaba (<0.3 $\mu$T) fall below the expected magnetic models by an order of magnitude. Assuming that accretion is magnetically-mediated and that the magnetic field decayed smoothly in the early solar system, the upper limits on the paleointensities of the CO and CV chondrites suggest an absence of nebula in the outer solar system nebula.  

$^{53}$Mn-$^{53}$Cr ages of fayalite from the CV3 Asuka 881317 indicate an alteration age of $4.2_{-0.7}^{+0.8}$ Myr after CAI formation \citep{doyle_early_2014}, which is within the error of the I-Xe age from Kaba of 4.50 $\pm$ 1.66 Myr after CAI formation \citep{pravdivtseva_i-xe_2013}. We note that the Kaba’s $^{53}$Mn-$^{53}$Cr age reported in \citep{gattacceca_new_2016} used the San Carlos olivine standard that needs to be corrected in future measurements \citep{doyle_early_2014}. Instead, we adopt the $^{53}$Mn-$^{53}$Cr ages of fayalite from the CV3 Asuka 881317 of $4.2_{-0.7}^{+0.8}$ Myr after CAI formation as the Kaba alteration age (Fig. \ref{fig:bulkfields}B). Together, the CV and CO magnetic data suggest the dissipation of the outer solar system solar nebula by 2.7-5 Myr after CAI-formation, with the CV magnetic record providing the most constrained evidence for the dissipation of the nebula in the outer solar system of $\sim$4.2 Myr after CAI-formation. This is broadly consistent with the recent models that suggest that chondrules from the CB carbonaceous chondrites Gujba and Hammadah al Hamra 237, which are thought to have formed, respectively, at $4.6_{-0.5}^{+0.5}$ Myr after CAI formation and $4.5_{-0.9}^{+0.9}$ after CAI formation as a result of high velocity impact between planetesimals in an environment where the nebular gas had already dissipated \citep{krot_young_2005} and with simulations of these impacts that indicate that the nebula dissipated by ~5 Myr after CAI formation \citep{johnson_timing_2016}. 

We can now establish the dissipation time of the solar nebula between the inner and outer solar system as well as within the outer solar system. Magnetic measurements suggest that the inner solar nebula dissipated by 3.71 $\pm$ 0.23 Myr after CAI formation \citep{wang_lifetime_2017, weiss_history_2021}. Comparing the angrite age with that of the CV chondrites would suggest that the solar nebula dissipated within <1.5 Myr between the two reservoirs. If, independently, we compare the CM chondrite age with that of the CV chondrites, we obtain a dispersal time in the outer solar system solar nebula of <1.6 Myr after the CM magnetic record. These two upper limits on the differential dissipation time between and within the reservoirs suggest a rapid dissipation of the disk: together with the evidence that the disk existed for at least $\sim$3 Ma, it overall supports a dual-timescale evolution of the solar system PPD. Models suggest that a two-timescale evolution of the disk is consistent with magnetically-driven and/or photoevaporative winds predominately driving PPD dissipation \citep{armitage_two_2013, clarke_dispersal_2001, shadmehri_time-dependent_2019} with potential influence of external photoevaporation aided by other stars \citep{concha-ramirez_external_2019}.

Finally, the dissipation of the disk sets the limit for the accretion of gas to the gas giants \citep{weiss_what_2021}. Thus, if we take the age of the CVs as the first evidence for the dissipation of the solar nebula in the outer solar system that indicates that the gas giants stopped accreting mass sometime between 3.5-5 Myr after CAI formation. 

\section{Future Directions for Solar Nebula Paleomagnetism}
While previous measurements \citep{cournede_early_2015, fu_fine-scale_2021, gattacceca_new_2016} and those reported in this study can potentially provide important information about the lifetime of the solar nebula, two key aspects that are fundamental to obtaining robust records of the solar nebula remain largely unconstrained. First, additional high-precision ages from meteorites are necessary. Determining when alteration occurred in various meteorite parent bodies is crucial to establish when CRMs may have been acquired. Future studies that include age measurements of the alteration products in the meteorites (ideally dating the formation of the ferromagnetic minerals of interest), including CO chondrites, will be able to provide the precision necessary to establish more precise timing of the solar nebula dispersal. 

Second, as briefly discussed in section 2.2, the CRM to ARM ratio for magnetite that formed during aqueous alteration of metal is undetermined. While chondrules and refractory inclusions likely acquired TRMs due to their pre-accretionary thermal histories, the magnetic record of bulk aqueously altered samples is usually carried in the form of CRMs. Unfortunately, very little is known about the mechanism of CRM acquisition during parent-body alteration and how it can affect the recovered paleointensity. A recent study suggests that magnetite can form through different pathways in chondrites \citep{sridhar_constraints_2021}. Magnetite grains in most CM and CO chondrites likely formed through pseudomorphic replacement of metal, while magnetite in framboids and plaquettes, which are found in comparatively high abundance inseveral chondrites including the CM Paris, are thought to have formed through the dissolution of sulfide and precipitation of magnetite in these exotic morphologies. For the case that magnetite forms through replacement of metal, previous studies of alteration pseudomorphic reactions in terrestrial samples have observed that certain minerals can inherent the NRM carried by the precursor phase \citep{heider_two_1987, jiang_remagnetization_2017}. Because FeNi metal across a bulk chondrite subsample is expected to carry a non-uniform remanence (because it will carry several non-unidirectional pre-accretionary remanences), it is possible that the absence of a remanence recorded by magnetite corresponds to a signal inherited from the metal. This could fundamentally impact the paleointensities obtained from CRMs carried by magnetite in CM chondrites \citep{cournede_early_2015}, CV chondrites \citep{fu_fine-scale_2021, gattacceca_new_2016} and CO chondrites in this study. For the CMs, for example, the lack of HT components that demagnetize between the Curie temperatures of pyrrhotite and magnetite in most CM chondrites from \cite{cournede_early_2015} was interpreted to be due to alteration during laboratory heating. A second interpretation is that the magnetite inherited the remanence from its parent metal and, as a result, does not contain a HT component associated with magnetite. This interpretation is consistent with pyrrhotite in CM chondrites carrying a CRM, indicating that a field was seemingly present at some point during the alteration sequence of these meteorites, but there being no record of this remanence in the magnetite among most meteorites in this group. Notably, the magnetite in Paris does appear to carry a CRM, which is consistent with this meteorite containing a large proportion of magnetite framboids and plaquettes that are likely to have recorded more traditional CRMs as the grains in these morphologies grew through their blocking volumes as they were precipitated. This interpretation could also explain the absence of a remanence in the CV Kaba at temperatures above the peak metamorphic temperature reached on the parent body \citep{gattacceca_new_2016}. This would suggest that the absence of a remanence may in fact not reflect the dissipation of the disk and its associated field, but instead the inability of magnetite to record a remanence when it forms through aqueous alteration of metal. To determine whether robust records of the solar nebula can in fact be recovered from bulk samples of aqueously altered chondrites, it is crucial to conduct experimental work to determine how magnetite in different families of chondrites and how they could acquire CRMs. 

\section{Conclusion}
We report paleomagnetic measurements of bulk matrix-rich samples from the CO chondrites ALHA77307 and DOM 08006. Assuming that magnetite could have recorded ambient fields during parent-body alteration, the CO chondrite records indicate that the ambient field was <0.9 $\mu$T. Combining these results with previous paleomagnetic measurements, we suggest that the nebular gas in the carbonaceous chondrite region ($\sim$3-7 AU) dissipated within 3.5-5 Myr after CAI formation. Comparing the time of the dissipation of the nebula in the outer solar system with that from the inner solar system indicates that the difference in dissipation ages between the two reservoirs was <1.5 Ma. This supports a dual-timescale evolution of the solar system PPD, consistent with magnetically driven winds and/or photoevaporation as processes that mediated the dissipation the solar system PPD. Additionally, the end of the solar nebula indicated that the gas giants stopped accreting by 3.5-5 Myr after CAI formation. We also highlight future work necessary for the field of solar nebula paleomagnetism which includes obtaining more ages from alteration products to obtain high-resolution temporal records from the nebula. We also highlight issues in using magnetite in bulk carbonaceous chondrites to determine nebula field records and the future work necessary for such records to be acquired, as well as obtaining more ages from alteration products to obtain high-resolution temporal records from the nebula. We also discuss how there is a possibility that paleomagnetic measurements of carbonaceous chondrites that have been interpreted as evidence for the absence of a field in the solar nebula, including this study, might be a result of non-ideal magnetic acquisition during parent-body alteration. Overall, future experimental work to understand the mechanisms of CRM acquisition in parent-bodies is critical to reliably obtain records of the solar nebula. 

\section*{Acknowledgments}
We thank E. Lima for helpful discussions. C.S.B. and B.P.W. thank the
NASA Discovery Program (contract NNM16AA09C) and Thomas F. Peterson, Jr. for support.

\section*{Data Availability Statement}
All data needed to evaluate the conclusions in the paper can be found through the Magnetics Information Consortium (MagIC) Database: doi:10.7288/V4/MAGIC/19362. 

\section*{Supplementary materials}
Supplementary text and materials are available at: \href{https://doi.org/10.1029/2021JE007139}{https://doi.org/10.1029/2021JE007139}.

\begin{landscape}
\begin{table}[htp]\centering
\footnotesize
\caption{\textbf{PCA fits for ALHA77307}. The first column shows the sample name, the second shows the weight (in mg) of the sample, the third shows the distance from the fusion crust (in mm; “fusion-crusted” if the sample contained fusion crust), the fourth shows the name of the component (LC = low coercivity, MC = medium coercivity, HC = high coercivity, LT = low temperature and HT = high temperature) with "*" indicating a non-magnetized range, the fifth shows the range of levels/temperatures used in the PCA fit (in mT or $^o$C), the sixth shows the number of data points in the PCA fit, the seventh, eighth and nineth show the declination, inclination and maximum angular deviation (MAD) for the fit without anchoring to the origin, the tenth shows the deviation angle (DANG) between anchored and non-origin-anchored fits, the eleventh, twelfth and thirteenth show the origin-anchored declination, origin-anchored inclination and origin-anchored MAD for the fits, and the fourteenth shows the NRM to ARM (AC field of 145 mT with a DC bias field of 50 $\mu$T) ratio. }\label{tab: }
\begin{adjustbox}{max width=\linewidth}
\scriptsize
\begin{tabular}{lccccccccccccc}\toprule
Sample &Weight (mg) &Distance from fusion crust (mm) &Component &Range (mT or °C) &N &Declination (°) &Inclination (°) &MAD (°) &DANG (°) &Anchored Declination (°) &Anchored Inclination (°) &Anchored MAD (°) &NRM/ARM \\\cmidrule{1-14}
ALHA.a &5.51 &fusion crusted &LC &0-20.5 mT &41 &118.5 &-4.9 &4.6 & & & & &\multirow{2}{*}{4.8} \\
& & &HC &26-410 mT &125 &288.6 &9.4 &3.4 &2.4 &289.8 &9.1 &2.7 & \\\cmidrule{1-14}
ALHA.b &7.68 &1.8 &LC &0-39 mT &64 &148.1 &11.5 &1.1 &2.5 &149 &12.1 &1.5 &\multirow{2}{*}{2.1} \\
& & &HC* &39-145 mT &87 &175.5 &3.7 &33.7 & & & & & \\\cmidrule{1-14}
ALHA.c &9.03 &3.2 &LC &0-58 mT &83 &155.5 &18.4 &1.8 &3.4 &154.7 &17.6 &2.2 &\multirow{2}{*}{2.2} \\
& & &HC* &58-145 mT &68 &358 &-1.3 &40.9 & & & & & \\\cmidrule{1-14}
ALHA.d &6.15 &4.9 &LC &0-55 mT &80 &163.2 &15.8 &3.6 &4.1 &161.7 &14.8 &3.4 &\multirow{2}{*}{1} \\
& & &HC* &55-145 mT &71 &157.5 &11.7 &38.6 & & & & & \\\cmidrule{1-14}
ALHA.e &1.1 &fusion crusted &LC &0-18.5 mT &36 &119.3 &-20.3 &3.9 & & & & & \\
& & &HT &251-580 °C &17 &277.4 &0 &10.9 &19.2 &290.7 &0.7 &10.7 & \\\cmidrule{1-14}
ALHA.f &0.6 &fusion crusted &LC &0-19 mT &39 &123.7 &-11 &5.1 & & & & & \\
& & &HT &251-580 °C &20 &281.3 &9.1 &16.1 &20.8 &294.7 &6.7 &10.5 & \\\cmidrule{1-14}
ALHA.g &2 &4.5 &LT &0-425 °C &11 &158.9 &5.6 &5.7 & & & & & \\
& & &HT* &425-580 °C &12 &357.1 &-3.3 &33.5 &34.9 &357.8 &27.7 &19.8 & \\\cmidrule{1-14}
ALHA.h &0.6 &5.8 &LC &0-21 mT &42 &152.8 &6.5 &5.5 & & & & & \\
& & &LT &51-375 °C &8 &112.9 &-2.1 &33.9 &32.9 &144.1 &-4.5 &14.1 & \\
& & &HT* &375-570 °C &13 &160.8 &-6.6 &30.5 & & & & & \\
\bottomrule
\end{tabular}
\end{adjustbox}
\label{bulkdirec1}
\end{table}
\end{landscape}

\begin{landscape}
\begin{table}[htp]\centering
\footnotesize
\caption{\textbf{PCA fits for DOM 08006}. The first column shows the sample name, the second shows the weight (in mg) of the sample, the third shows the distance from the fusion crust (in mm; “fusion-crusted” if the sample contained fusion crust), the fourth shows the name of the component (LC = low coercivity, MC = medium coercivity, HC = high coercivity, LT = low temperature and HT = high temperature), the fifth shows the range of levels/temperatures used in the PCA fit (in mT or $^o$C), the sixth shows the number of data points in the PCA fit, the seventh, eighth and nineth show the declination, inclination and maximum angular deviation (MAD) for the fit without anchoring to the origin, the tenth shows the deviation angle (DANG) between anchored and non-origin-anchored fits, the eleventh, twelfth and thirteenth show the origin-anchored declination, origin-anchored inclination and origin-anchored MAD for the fits, and the fourteenth shows the NRM to ARM (AC field of 145 mT with a DC bias field of 50 $\mu$T) ratio. }\label{tab: }
\begin{adjustbox}{max width=\linewidth}
\scriptsize
\begin{tabular}{lccccccccccccc}\toprule
Sample &Weight (mg) &Distance from fusion crust (mm) &Component &Range (mT or °C) &N &Declination (°) &Inclination (°) &MAD (°) &DANG (°) &Anchored Declination (°) &Anchored Inclination (°) &Anchored MAD (°) &NRM/ARM \\\cmidrule{1-14}
DOM.a &1.03 &fusion crusted &HC &0-145 mT &150 &32.8 &46.9 &2.2 &1.6 &32.6 &47.7 &1.8 &2.75 \\\cmidrule{1-14}
DOM.b &1.37 &1 &LC &0-10 mT &20 &30.4 &3.9 &4 &12.6 &30.4 &-4 &6.6 &\multirow{2}{*}{0.14} \\
& & &HC* &10-145 mT &131 &147.7 &-77.1 &40.3 & & & & & \\\cmidrule{1-14}
DOM.c &2.34 &2 &LC &0-7.5 mT &15 &29.6 &6.7 &3.7 & & & & &\multirow{2}{*}{0.04} \\
& & &HC &19-145 mT &113 &236.3 &-37.6 &17.3 &13.3 &244.9 &-36.3 &13.9 & \\\cmidrule{1-14}
DOM.d &1.61 &3.1 &LC &0-6 mT &12 &34.3 &16.8 &5.5 & & & & &\multirow{4}{*}{0.27} \\
& & &MC &6-17.5 mT &24 &80.7 &-30.2 &30 & & & & & \\
& & &MC &17.5-32 mT &23 &204.5 &-62.6 &25.6 &48.9 &303.4 &-54.7 &13.4 & \\
& & &HC* &32-145 mT &94 &309.1 &11 &39 & & & & & \\\cmidrule{1-14}
DOM.j &1.75 &7.6 &LC &0-28 mT &53 &120.2 &53.4 &4.4 & & & & &\multirow{3}{*}{0.64} \\
& & &MC &28-45 mT &18 &200.8 &75.4 &18.8 &7.9 &189.8 &69.9 &9.2 & \\
& & &HC* &45-145 mT &81 &149.7 &38.7 &49 & & & & & \\\cmidrule{1-14}
DOM.k &3.28 &6.9 &LC &0-9.5 mT &19 &66.1 &22.6 &4.9 & & & & &\multirow{3}{*}{0.24} \\
& & &MC &9.5-24 mT &30 &72.9 &46.5 &12.8 &39.5 &114 &32.1 &14.8 & \\
& & &HC* &24-145 mT &103 &171.8 &-30.6 &42.3 & & & & & \\\cmidrule{1-14}
DOM.l &3.5 &5.8 &LC &0-6 mT &12 &35.6 &11.7 &2.6 & & & & &\multirow{3}{*}{0.33} \\
& & &MC (?) &6-28 mT &42 &207 &12.9 &40.3 & & & & & \\
& & &HC &28-145 mT &98 &100.4 &-22.7 &35.2 &11.3 &94.7 &-15.5 &21 & \\\cmidrule{1-14}
DOM.m &1.71 &4.8 &LC &1-15.5 mT &30 &45 &29.7 &10.6 &37.9 &57.5 &50.1 &19.8 &\multirow{2}{*}{0.22} \\
& & &HC* &15.5-145 mT &120 &119.8 &54.2 &44.4 & & & & & \\\cmidrule{1-14}
DOM.g &1.78 &2 &0-350 °C &8 &357.1 &-23.2 &20.2 &10 &356.8 &-32.8 &5 & \\
& & &HT* &350-650 °C &7 &352.2 &-37.9 &33.8 & & & & & \\\cmidrule{1-14}
DOM.ha &1.13 &3 &0-350 °C &8 &39.2 &-9.1 &8.2 & & & & & \\
& & &HT* &350-650 °C &7 &10.6 &15.9 &33.4 & & & & & \\\cmidrule{1-14}
DOM.hb &2.23 &3.1 &0-200 °C &5 &42.3 &-9 &7.4 & & & & & \\
& & &HT* &200-650 °C &10 &335.6 &30.2 &40.2 & & & & & \\\cmidrule{1-14}
DOM.ia &2.71 &3.8 &0-350 °C &8 &34.8 &-6.4 &13.9 & & & & & \\
& & &HT* &400-600 °C &5 &198.8 &-6.1 &20.1 &13.8 &187.8 &-2.4 &10.1 & \\\cmidrule{1-14}
DOM.ib &0.67 &3.8 &0-350 °C &8 &34.8 &-6.4 &13.9 & & & & & \\
& & &HT* &400-600 °C &5 &171.5 &8.8 &27.1 &10.6 &179.4 &12.8 &13.7 & \\
\bottomrule
\end{tabular}
\end{adjustbox}
\label{bulkdirec2}
\end{table}
\end{landscape}
\begin{table}[!htp]\centering
\footnotesize
\caption{\textbf{Paleointensities from the ARM experiment with samples from ALHA77307 and DOM 08006}. The first column shows the sample name, the second shows the distance from the fusion crust (in mm; “fusion-crusted” if the sample contained fusion crust), the third shows the range of AF levels used in the paleointensity fit (in mT), the fourth shows the number of datapoints used in the paleointensity fit, the fifth shows the parametric correlation ($\rho$) between NRM and ARM datasets, the sixth shows the type of fit used to calculate the paleointensity (OLS = ordinary least square; RMA = reduced major axis), the seventh shows the paleointensity (in $\mu$T), the eighth shows the calculated 95\% confidence interval of the paleointensity fit (in $\mu$T) and the nineth shows the recording limit of the sample (in $\mu$T). For each sample (ALHA77307 and DOM 08006) we calculated mean paleointensities and the 95\% confidence interval.}\label{tab: }
\begin{adjustbox}{max width=\linewidth}
\scriptsize
\begin{tabular}{lccccccccccccc}\toprule
Sample &Distance from fusion crust (mm) &Range (mT) &N &$\rho$ &Fit Type &Paleofield ($\mu$T) &95\% conf. int. ($\mu$T) &Recording limit ($\mu$T) \\\cmidrule{1-9}
ALHA.a &fusion crusted &0-10 &20 &1 &OLS &358.5 &15.8 &- \\
& &26-145 &100 &0.99 &RMA &26.7 &0.8 &- \\\cmidrule{1-9}
ALHA.b &7.68 &0-10 &20 &0.99 &OLS &206.9 &10.2 &- \\
& &39-145 &87 &0.8 &RMA &2 &0.3 &- \\\cmidrule{1-9}
ALHA.c &3.2 &0-9 &18 &0.99 &OLS &247.7 &16.3 &- \\
& &55-145 &71 &0.5 &OLS &1.5 &0.5 &> 1.5 \\\cmidrule{1-9}
ALHA.d &4.9 &0-10 &20 &0.99 &OLS &113.3 &7.3 &- \\
& &55-145 &71 &0.54 &OLS &3 &0.6 &> 0.9 \\\midrule
& & & & &\textbf{Mean*} &\textbf{2.2} &\multicolumn{2}{c}{*only high coercvity and no fusion crust} \\
& & & & &\textbf{95\% conf. int} &\textbf{0.9} &\textbf{} & \\
& & & & &\textbf{} &\textbf{} &\textbf{} & \\\cmidrule{1-9}
DOM.a &fusion crusted &0-145 &150 &1 &RMA &39.5 &0.4 &- \\\cmidrule{1-9}
DOM.b &0.9 &0-10 &20 &0.97 &OLS &16.4 &2.2 &- \\
& &10-145 &131 &0.14 &OLS &0.1 &0.1 &> 1.5 \\\cmidrule{1-9}
DOM.c &2 &0-7.5 &15 &0.98 &OLS &28 &3.4 &- \\
& &19-145 &113 &0.97 &RMA &4.1 &0.2 &> 0.3 \\\cmidrule{1-9}
DOM.d &3.1 &0-6 &12 &0.93 &OLS &29.5 &8.1 &- \\
& &6-17.5 &24 &0.94 &RMA &5.4 &0.8 &- \\
& &17.5-32 &23 &0.87 &RMA &6.4 &1.5 &- \\
& &32-145 &94 &- &OLS &-0.02 &0.7 &> 3 \\\cmidrule{1-9}
DOM.j &7.6 &0-28 &53 &1 &RMA &18.5 &0.6 &- \\
& &28-45 &18 &1 &OLS &7.1 &1.8 &- \\
& &45-145 &81 &0.27 &OLS &0.8 &0.6 &> 0.9 \\\cmidrule{1-9}
DOM.k &6.9 &0-9.5 &19 &1 &OLS &17.9 &1.3 &- \\
& &9.5-24 &30 &0.97 &RMA &5.7 &0.5 &- \\
& &24-145 &103 &- &OLS &-0.3 &0.2 &> 1.5 \\\cmidrule{1-9}
DOM.l &5.8 &0-6 &12 &0.97 &OLS &34.1 &5.7 &- \\
& &28-145 &98 &0.83 &RMA &3.4 &0.4 &> 0.9 \\\cmidrule{1-9}
DOM.m &4.8 &1-15.5 &30 &0.94 &RMA &12.9 &1.6 &- \\
& &15.5-145 &120 &0.3 &OLS &0.3 &0.2 &> 1.5 \\\cmidrule{1-9}
& & & & &\textbf{Mean**} &\textbf{1.2} &\multicolumn{2}{c}{**only high coercvity and no fusion crust} \\
& & & & &\textbf{95\% conf. int} &\textbf{1.4} & & \\\cmidrule{6-9}
& & & & &\textbf{Mean***} &\textbf{0.2} &\multicolumn{2}{c}{***only high coercvity, no fusion crust, DOM.c and DOM.l not included} \\
& & & & &\textbf{95\% conf. int} &\textbf{0.4} & & \\
\bottomrule
\end{tabular}
\end{adjustbox}
\label{bulkARM}
\end{table}
\begin{landscape}
\begin{table}[!htp]\centering
\footnotesize
\caption{\textbf{Paleointensities and calculated paleomagnetic criteria [PICRIT03 from \cite{paterson_improving_2014}] from the IZZI experiment for ALHA77307 and DOM 08006}. The first column shows the sample name, the second shows the distance from the fusion crust (in mm; “fusion-crusted” if the sample contained fusion crust), the third shows the range of temperatures used in the paleointensity fit (in $^o$C), the fourth shows the number of datapoints used in the fit, the fifth shows the paleointensity (in $\mu$T), the sixth shows the calculated 95\% confidence interval of the paleointensity fit (in $\mu$T), the seventh, eighth, nineth, tenth, eleventh and twelfth show the paleomagnetic parameters DRATS, $f$, $\beta$, $q$, CDRAT and maximum DRAT from \cite{paterson_improving_2014}, and the thirteenth shows the number of partial TRM (pTRM) checks conducted during the IZZI experiment.}\label{tab: }
\begin{adjustbox}{max width=\linewidth}
\scriptsize
\begin{tabular}{lcccccccccccccc}\toprule
Sample &Distance from fusion crust (mm) &Range ($^o$C) &$N$ &Lab field direction (dec,inc) (°) &Paleofield ($\mu$T) &95\% conf int ($\mu$T) &DRATS &$f$ &$\beta$ &$q$ &CDRAT &max. DRAT &number of pTRM checks \\\cmidrule{1-14}
ALHA.e &fusion crust &400-580 &13 &(360,0) &13.9 &4.7 &5 &1.1 &0.3 &2.8 &5.1 &17.8 &6 \\\cmidrule{1-14}
ALHA.f &fusion crust &400-580 &13 &(360,0) &13.9 &5.4 &0.5 &1 &0.4 &2.3 &0.5 &7.8 &6 \\\cmidrule{1-14}
ALHA.g &4.5 &540-580 &12 &(360,0) &1.5 &8.9 &34.6 &0.4 &5.9 &-0.1 &51.9 &23.4 &6 \\\cmidrule{1-14}
ALHA.h &5.8 &475-580 &10 &(360,0) &-7.3 &10.2 &3 &1.5 &1.4 &-0.8 &3.9 &62.8 &6 \\\cmidrule{1-14}
DOM.ia &3.8 &0-350 &8 &(180,0) &23.1 &20.7 &35 &2.5 &0.9 &2.2 &30.4 &17.1 &3 \\
& &400-600 &5 & &2.8 &4.5 &13 &0.7 &1.6 &0.2 &14.9 &4.9 &5 \\\cmidrule{1-14}
DOM.ib &3.8 &0-350 &8 &(180,0) &23.1 &20.7 &35 &2.5 &0.9 &2.2 &30.4 &17.1 &3 \\
& &400-600 &5 & &3.6 &5.1 &13 &0.9 &1.4 &0.4 &14.7 &5.3 &5 \\\cmidrule{1-14}
\end{tabular}
\end{adjustbox}
\label{bulkTHERMAL}
\end{table}
\end{landscape}

\section{References}
\bibliographystyle{apalike}
\begingroup
\renewcommand{\section}[2]{}
\bibliography{references}

\begin{thebibliography}{}

\bibitem[Alexander et~al., 2018]{alexander_mutli-technique_2018}
Alexander, C. M.~O., Greenwood, R.~C., Bowden, R., Gibson, J.~M., Howard,
  K.~T., and Franchi, I.~A. (2018).
\newblock A mutli-technique search for the most primitive {CO} chondrites.
\newblock {\em Geochimica et Cosmochimica Acta}, 221:406--420.

\bibitem[Alexander et~al., 2006]{alexander_photoevaporation_2006}
Alexander, R.~D., Clarke, C.~J., and Pringle, J.~E. (2006).
\newblock Photoevaporation of protoplanetary discs – {II}. {Evolutionary}
  models and observable properties.
\newblock {\em Monthly Notices of the Royal Astronomical Society},
  369(1):229--239.

\bibitem[Armitage and Kley, 2019]{audard_protoplanetary_2019}
Armitage, P.~J. and Kley, W. (2019).
\newblock {\em From {Protoplanetary} {Disks} to {Planet} {Formation}}.
\newblock Saas-{Fee} {Advanced} {Course}. Springer-Verlag, Berlin Heidelberg.

\bibitem[Armitage et~al., 2013]{armitage_two_2013}
Armitage, P.~J., Simon, J.~B., and Martin, R.~G. (2013).
\newblock Two timescale dispersal of magnetized protplanetary disks.
\newblock {\em The Astrophysical Journal}, 778(1):L14.
\newblock Publisher: IOP Publishing.

\bibitem[Bai, 2016]{bai_toward_2016}
Bai, X. (2016).
\newblock Toward a global evolutionary model of protoplanetary disks.
\newblock {\em Astrophys. J.}, 821:80.

\bibitem[Bai and Goodman, 2009]{bai_heat_2009}
Bai, X.-N. and Goodman, J. (2009).
\newblock Heat and dust in active layers of protostellar disks.
\newblock {\em The Astrophysical Journal}, 701(1):737--755.

\bibitem[Biersteker et~al., 2019]{biersteker_implications_2019}
Biersteker, J.~B., Weiss, B.~P., Heinisch, P., Herčik, D., Glassmeier, K.-H.,
  and Auster, H.-U. (2019).
\newblock Implications of {Philae} magnetometry {Measurements} at comet
  {67P}/{Churyumov}–{Gerasimenko} for the nebular field of the outer solar
  system.
\newblock {\em The Astrophysical Journal}, 875(1):39.
\newblock Publisher: American Astronomical Society.

\bibitem[Bonal et~al., 2007]{bonal_organic_2007}
Bonal, L., Bourot-Denise, M., Quirico, E., Montagnac, G., and Lewin, E. (2007).
\newblock Organic matter and metamorphic history of {CO} chondrites.
\newblock {\em Geochimica et Cosmochimica Acta}, 71(6):1605--1623.

\bibitem[Bonal et~al., 2016]{bonal_thermal_2016}
Bonal, L., Quirico, E., Flandinet, L., and Montagnac, G. (2016).
\newblock Thermal history of type 3 chondrites from the {Antarctic} meteorite
  collection determined by {Raman} spectroscopy of their polyaromatic
  carbonaceous matter.
\newblock {\em Geochimica et Cosmochimica Acta}, 189:312--337.

\bibitem[Borlina et~al., 2021]{borlina_paleomagnetic_2021}
Borlina, C.~S., Weiss, B.~P., Bryson, J. F.~J., Bai, X.-N., Lima, E.~A.,
  Chatterjee, N., and Mansbach, E.~N. (2021).
\newblock Paleomagnetic evidence for a disk substructure in the early solar
  system.
\newblock {\em Science Advances}, 7(42):eabj6928.
\newblock Publisher: American Association for the Advancement of Science.

\bibitem[Brasser and Mojzsis, 2020]{brasser_partitioning_2020}
Brasser, R. and Mojzsis, S.~J. (2020).
\newblock The partitioning of the inner and outer {Solar} {System} by a
  structured protoplanetary disk.
\newblock {\em Nature Astronomy}, 4(5):492--499.
\newblock Number: 5 Publisher: Nature Publishing Group.

\bibitem[Bryson et~al., 2019]{bryson_constraints_2019}
Bryson, J. F.~J., Neufeld, J.~A., and Nimmo, F. (2019).
\newblock Constraints on asteroid magnetic field evolution and the radii of
  meteorite parent bodies from thermal modelling.
\newblock {\em Earth and Planetary Science Letters}, 521:68--78.

\bibitem[Bryson et~al., 2017]{bryson_paleomagnetic_2017}
Bryson, J. F.~J., Weiss, B.~P., Harrison, R.~J., Herrero-Albillos, J., and
  Kronast, F. (2017).
\newblock Paleomagnetic evidence for dynamo activity driven by inward
  crystallisation of a metallic asteroid.
\newblock {\em Earth and Planetary Science Letters}, 472:152--163.

\bibitem[Bryson et~al., 2020]{bryson_evidence_2020}
Bryson, J. F.~J., Weiss, B.~P., Lima, E.~A., Gattacceca, J., and Cassata, W.~S.
  (2020).
\newblock Evidence for asteroid scattering and distal solar system solids from
  meteorite paleomagnetism.
\newblock {\em The Astrophysical Journal}, 892(2):126.
\newblock Publisher: American Astronomical Society.

\bibitem[Clarke et~al., 2001]{clarke_dispersal_2001}
Clarke, C.~J., Gendrin, A., and Sotomayor, M. (2001).
\newblock The dispersal of circumstellar discs: the role of the ultraviolet
  switch.
\newblock {\em Monthly Notices of the Royal Astronomical Society},
  328(2):485--491.

\bibitem[Concha-Ramírez et~al., 2019]{concha-ramirez_external_2019}
Concha-Ramírez, F., Wilhelm, M. J.~C., Portegies Zwart, S., and Haworth,
  T.~J. (2019).
\newblock External photoevaporation of circumstellar discs constrains the
  time-scale for planet formation.
\newblock {\em Monthly Notices of the Royal Astronomical Society},
  490(4):5678--5690.

\bibitem[Cournede et~al., 2015]{cournede_early_2015}
Cournede, C., Gattacceca, J., Gounelle, M., Rochette, P., Weiss, B.~P., and
  Zanda, B. (2015).
\newblock An early solar system magnetic field recorded in {CM} chondrites.
\newblock {\em Earth and Planetary Science Letters}, 410:62--74.

\bibitem[Davidson et~al., 2019]{davidson_mineralogy_2019}
Davidson, J., Alexander, C. M.~O., Stroud, R.~M., Busemann, H., and Nittler,
  L.~R. (2019).
\newblock Mineralogy and petrology of {Dominion} {Range} 08006: {A} very
  primitive {CO3} carbonaceous chondrite.
\newblock {\em Geochimica et Cosmochimica Acta}, 265:259--278.

\bibitem[DeMeo and Carry, 2014]{demeo_solar_2014}
DeMeo, F.~E. and Carry, B. (2014).
\newblock Solar {System} evolution from compositional mapping of the asteroid
  belt.
\newblock {\em Nature}, 505(7485):629--634.
\newblock Number: 7485 Publisher: Nature Publishing Group.

\bibitem[Desch et~al., 2018]{desch_effect_2018}
Desch, S.~J., Kalyaan, A., and Alexander, C. M.~O. (2018).
\newblock The effect of {Jupiter}'s formation on the distribution of refractory
  elements and inclusions in meteorites.
\newblock {\em The Astrophysical Journal Supplement Series}, 238(1):11.

\bibitem[Doyle et~al., 2014]{doyle_early_2014}
Doyle, P.~M., Jogo, K., Nagashima, K., Krot, A.~N., Wakita, S., Ciesla, F.~J.,
  and Hutcheon, I.~D. (2014).
\newblock Early aqueous activity on the ordinary and carbonaceous chondrite
  parent bodies recorded by fayalite.
\newblock {\em Nature Commun.}, 6:7444.

\bibitem[Dullemond and Dominik, 2005]{dullemond_dust_2005}
Dullemond, C.~P. and Dominik, C. (2005).
\newblock Dust coagulation in protoplanetary disks: {A} rapid depletion of
  small grains.
\newblock {\em Astron. Astrophys.}, 434:971--986.

\bibitem[Dunlop, 2002]{dunlop_theory_2002}
Dunlop, D.~J. (2002).
\newblock Theory and application of the {Day} plot ({Mrs}/{Ms} versus
  {Hcr}/{Hc}) - 2. {Application} to data for rocks, sediments, and soils.
\newblock {\em J. Geophys. Res.}, 107:EPM 5--1--EPM 5--15.

\bibitem[Dunlop and Özdemir, 1997]{dunlop_rock_1997}
Dunlop, D.~J. and Özdemir, O. (1997).
\newblock {\em Rock {Magnetism}: {Fundamentals} and {Frontiers}}.
\newblock Cambridge {Studies} in {Magnetism}. Cambridge University Press, New
  York.

\bibitem[Dyl et~al., 2012]{dyl_early_2012}
Dyl, K.~A., Bischoff, A., Ziegler, K., Young, E.~D., Wimmer, K., and Bland,
  P.~A. (2012).
\newblock Early solar system hydrothermal activity in chondritic asteroids on
  1–10-year timescales.
\newblock {\em Proceedings of the National Academy of Sciences},
  109(45):18306--18311.
\newblock Publisher: National Academy of Sciences Section: Physical Sciences.

\bibitem[Ercolano and Pascucci, 2017]{ercolano_dispersal_2017}
Ercolano, B. and Pascucci, I. (2017).
\newblock The dispersal of planet-forming discs: theory confronts observations.
\newblock {\em Royal Society Open Science}, 4(4):170114.
\newblock Publisher: Royal Society.

\bibitem[Fu et~al., 2020]{fu_weak_2020}
Fu, R.~R., Kehayias, P., Weiss, B.~P., Schrader, D.~L., Bai, X.-N., and Simon,
  J.~B. (2020).
\newblock Weak magnetic fields in the outer solar nebula recorded in {CR}
  chondrites.
\newblock {\em Journal of Geophysical Research: Planets}, (5):e2019JE006260.
\newblock \_eprint:
  https://agupubs.onlinelibrary.wiley.com/doi/pdf/10.1029/2019JE006260.

\bibitem[Fu et~al., 2021]{fu_fine-scale_2021}
Fu, R.~R., Volk, M. W.~R., Bilardello, D., Libourel, G., Lesur, G. R.~J., and
  Dor, O.~B. (2021).
\newblock The fine-scale magnetic history of the {Allende} meteorite:
  implications for the structure of the solar nebula.
\newblock {\em AGU Advances}, 2(3):e2021AV000486.
\newblock \_eprint:
  https://agupubs.onlinelibrary.wiley.com/doi/pdf/10.1029/2021AV000486.

\bibitem[Fu et~al., 2014]{fu_solar_2014}
Fu, R.~R., Weiss, B.~P., Lima, E.~A., Harrison, R.~J., Bai, X.-N., Desch,
  S.~J., Ebel, D.~S., Suavet, C., Wang, H., Glenn, D., Le~Sage, D., Kasama, T.,
  Walsworth, R.~L., and Kuan, A.~T. (2014).
\newblock Solar nebula magnetic fields recorded in the {Semarkona} meteorite.
\newblock {\em Science}, 346:1089--1092.

\bibitem[Fujiya et~al., 2012]{fujiya_evidence_2012}
Fujiya, W., Sugiura, N., Hotta, H., Ichimura, K., and Sano, Y. (2012).
\newblock Evidence for the late formation of hydrous asteroids from young
  meteoritic carbonates.
\newblock {\em Nature Communications}, 3(1):627.
\newblock Number: 1 Publisher: Nature Publishing Group.

\bibitem[Ganino and Libourel, 2020]{ganino_fumarolic-like_2020}
Ganino, C. and Libourel, G. (2020).
\newblock Fumarolic-like activity on carbonaceous chondrite parent body.
\newblock {\em Science Advances}, 6(27):eabb1166.
\newblock Publisher: American Association for the Advancement of Science.

\bibitem[Gattacceca and Rochette, 2004]{gattacceca_toward_2004}
Gattacceca, J. and Rochette, P. (2004).
\newblock Toward a robust normalized magnetic paleointensity method applied to
  meteorites.
\newblock {\em Earth and Planetary Science Letters}, 227(3):377--393.

\bibitem[Gattacceca et~al., 2016]{gattacceca_new_2016}
Gattacceca, J., Weiss, B.~P., and Gounelle, M. (2016).
\newblock New constraints on the magnetic history of the {CV} parent body and
  the solar nebula from the {Kaba} meteorite.
\newblock {\em Earth Planet. Sci. Lett.}, 455:166--175.

\bibitem[Gorti et~al., 2016]{gorti_disk_2016}
Gorti, U., Liseau, R., Sándor, Z., and Clarke, C. (2016).
\newblock Disk dispersal: theoretical understanding and observational
  constraints.
\newblock {\em Space Science Reviews}, 205(1):125--152.

\bibitem[Grossman and Brearley, 2005]{grossman_onset_2005}
Grossman, J.~N. and Brearley, A.~J. (2005).
\newblock The onset of metamorphism in ordinary and carbonaceous chondrites.
\newblock {\em Meteoritics \& Planetary Science}, 40(1):87--122.

\bibitem[Haisch et~al., 2001]{haisch_disk_2001}
Haisch, K.~E., Lada, E.~A., and Lada, C.~J. (2001).
\newblock Disk frequencies and lifetimes in young clusters.
\newblock {\em Astrophys. J.}, 553:L153--L156.

\bibitem[Hartmann et~al., 1998]{hartmann_accretion_1998}
Hartmann, L., Calvet, N., Gullbring, E., and D'Alessio, P. (1998).
\newblock Accretion and the {Evolution} of {T} {Tauri} {Disks}.
\newblock {\em Astrophys. J.}, 495(1):385.
\newblock Publisher: IOP Publishing.

\bibitem[Heider and Dunlop, 1987]{heider_two_1987}
Heider, F. and Dunlop, D.~J. (1987).
\newblock Two types of chemical remanent magnetization during the oxidation of
  magnetite.
\newblock {\em Physics of the Earth and Planetary Interiors}, 46(1):24--45.

\bibitem[Herndon et~al., 1976]{herndon_thermomagnetic_1976}
Herndon, J.~M., Rowe, M.~W., Larson, E.~E., and Watson, D.~E. (1976).
\newblock Thermomagnetic analysis of meteorites, 3. {C3} and {C4} chondrites.
\newblock {\em Earth and Planetary Science Letters}, 29(2):283--290.

\bibitem[Hernández et~al., 2007]{hernandez_spitzer_2007}
Hernández, J., Hartmann, L., Megeath, T., Gutermuth, R., Muzerolle, J.,
  Calvet, N., Vivas, A.~K., Briceño, C., Allen, L., Stauffer, J., Young, E.,
  and Fazio, G. (2007).
\newblock A {Spitzer} {Space} {Telescope} study of disks in the young σ
  {Orionis} cluster.
\newblock {\em The Astrophysical Journal}, 662:1067--1081.
\newblock ADS Bibcode: 2007ApJ...662.1067H.

\bibitem[Ireland and Kraus, 2008]{ireland_disk_2008}
Ireland, M.~J. and Kraus, A.~L. (2008).
\newblock The disk around {CoKu} {Tauri}/4: circumbinary, not transitional.
\newblock {\em The Astrophysical Journal}, 678(1):L59.
\newblock Publisher: IOP Publishing.

\bibitem[Jiang et~al., 2017]{jiang_remagnetization_2017}
Jiang, Z., Liu, Q., Dekkers, M.~J., Zhao, X., Roberts, A.~P., Yang, Z., Jin,
  C., and Liu, J. (2017).
\newblock Remagnetization mechanisms in {Triassic} red beds from {South}
  {China}.
\newblock {\em Earth and Planetary Science Letters}, 479:219--230.

\bibitem[Johnson et~al., 2016]{johnson_timing_2016}
Johnson, B.~C., Walsh, K.~J., Minton, D.~A., Krot, A.~N., and Levison, H.~F.
  (2016).
\newblock Timing of the formation and migration of giant planets as constrained
  by {CB} chondrites.
\newblock {\em Science Advances}, 2(12):e1601658.
\newblock Publisher: American Association for the Advancement of Science
  Section: Research Article.

\bibitem[Kirschvink, 1980]{kirschvink_least-squares_1980}
Kirschvink, J.~L. (1980).
\newblock The least-squares line and plane and the analysis of paleomagnetic
  data: examples from {Siberia} and {Morocco}.
\newblock {\em Geophys. J. R. Astr. Soc.}, 62:699--718.

\bibitem[Kirschvink et~al., 2008]{kirschvink_rapid_2008}
Kirschvink, J.~L., Kopp, R.~E., Raub, T.~D., Baumgartner, C.~T., and Holt,
  J.~W. (2008).
\newblock Rapid, precise, and high-sensitivity acquisition of paleomagnetic and
  rock-magnetic data: {Development} of a low-noise automatic sample changing
  system for superconducting rock magnetometers.
\newblock {\em Geochemistry, Geophysics, Geosystems}, 9(5).
\newblock \_eprint:
  https://agupubs.onlinelibrary.wiley.com/doi/pdf/10.1029/2007GC001856.

\bibitem[Kita and Ushikubo, 2011]{kita_evolution_2011}
Kita, N.~T. and Ushikubo, T. (2011).
\newblock Evolution of protoplanetary disk inferred from $^{\textrm{26}}${Al}
  chronology of individual chondrules.
\newblock {\em Meteorit. Planet. Sci.}, 47:1108--1119.

\bibitem[Kletetschka and Wieczorek, 2017]{kletetschka_fundamental_2017}
Kletetschka, G. and Wieczorek, M.~A. (2017).
\newblock Fundamental relations of mineral specific magnetic carriers for
  paleointensity determination.
\newblock {\em Physics of the Earth and Planetary Interiors}, 272:44--49.

\bibitem[Krot et~al., 2005]{krot_young_2005}
Krot, A.~N., Amelin, Y., Cassen, P., and Meibom, A. (2005).
\newblock Young chondrules in {CB} chondrites from a giant impact in the early
  {Solar} {System}.
\newblock {\em Nature}, 436(7053):989--992.
\newblock Number: 7053 Publisher: Nature Publishing Group.

\bibitem[Krot et~al., 2006]{krot_timescales_2006}
Krot, A.~N., Hutcheon, I.~D., Brearley, A.~J., Pravdivtseva, O.~V., Petaev,
  M.~I., and Hohenberg, C.~M. (2006).
\newblock Timescales and settings for alteration of chondritic meteorites.
\newblock In Lauretta, D.~S. and McSween, H.~Y., editors, {\em Meteorites and
  the {Early} {Solar} {System} {II}}, pages 525--553. University of Arizona
  Press, Tucson.

\bibitem[Kruijer et~al., 2020]{kruijer_great_2020}
Kruijer, T.~S., Kleine, T., and Borg, L.~E. (2020).
\newblock The great isotopic dichotomy of the early {Solar} {System}.
\newblock {\em Nature Astronomy}, 4(1):32--40.
\newblock Number: 1 Publisher: Nature Publishing Group.

\bibitem[Li et~al., 2021]{li_formation_2021}
Li, Y., Rubin, A.~E., and Hsu, W. (2021).
\newblock Formation of metallic-{Cu}-bearing mineral assemblages in type-3
  ordinary and {CO} chondrites.
\newblock {\em American Mineralogist}, 106(11):1751--1767.

\bibitem[McClelland, 1996]{mcclelland_theory_1996}
McClelland, E. (1996).
\newblock Theory of {CRM} acquired by grain growth, and its implications for
  {TRM} discrimination and palaeointensity determination in igneous rocks.
\newblock {\em Geophysical Journal International}, 126(1):271--280.
\newblock Publisher: Oxford Academic.

\bibitem[Mighani et~al., 2020]{mighani_end_2020}
Mighani, S., Wang, H., Shuster, D.~L., Borlina, C.~S., Nichols, C. I.~O., and
  Weiss, B.~P. (2020).
\newblock The end of the lunar dynamo.
\newblock {\em Science Advances}, 6(1):eaax0883.

\bibitem[Morbidelli et~al., 2015]{morbidelli_dynamical_2015}
Morbidelli, A., Walsh, K.~J., O'Brien, D.~P., Minton, D.~A., and Bottke, W.~F.
  (2015).
\newblock The dynamical evolution of the asteroid belt.
\newblock {\em Asteroids IV}, pages 493--507.

\bibitem[Nagata et~al., 1991]{nagata_magnetic_1991}
Nagata, T., Funaki, M., and Kojima, H. (1991).
\newblock Magnetic properties and natural remanent magnetization of
  carbonaceous chondrites containing pyrrhotite.
\newblock {\em Antarctic Meteorite Research}, 4:390.

\bibitem[Nagy et~al., 2019]{nagy_thermomagnetic_2019}
Nagy, L., Williams, W., Tauxe, L., Muxworthy, A.~R., and Ferreira, I. (2019).
\newblock Thermomagnetic recording fidelity of nanometer-sized iron and
  implications for planetary magnetism.
\newblock {\em Proc. Natl. Acad. Sci. USA}, 116(6):1984--1991.

\bibitem[Oran et~al., 2018]{oran_were_2018}
Oran, R., Weiss, B.~P., and Cohen, O. (2018).
\newblock Were chondrites magnetized by the early solar wind?
\newblock {\em Earth and Planetary Science Letters}, 492:222--231.

\bibitem[Owen, 2016]{owen_origin_2016}
Owen, J.~E. (2016).
\newblock The origin and evolution of transition discs: successes, problems,
  and open questions.
\newblock {\em Publications of the Astronomical Society of Australia}, 33.
\newblock Publisher: Cambridge University Press.

\bibitem[O’Brien et~al., 2020]{obrien_arrival_2020}
O’Brien, T., Tarduno, J.~A., Anand, A., Smirnov, A.~V., Blackman, E.~G.,
  Carroll-Nellenback, J., and Krot, A.~N. (2020).
\newblock Arrival and magnetization of carbonaceous chondrites in the asteroid
  belt before 4562 million years ago.
\newblock {\em Communications Earth \& Environment}, 1(1):1--7.
\newblock Bandiera\_abtest: a Cc\_license\_type: cc\_by Cg\_type: Nature
  Research Journals Number: 1 Primary\_atype: Research Publisher: Nature
  Publishing Group Subject\_term: Asteroids, comets and Kuiper belt;Early solar
  system;Meteoritics Subject\_term\_id:
  asteroids-comets-and-kuiper-belt;early-solar-system;meteoritics.

\bibitem[Paterson et~al., 2014]{paterson_improving_2014}
Paterson, G.~A., Tauxe, L., Biggin, A.~J., Shaar, R., and Jonestrask, L.~C.
  (2014).
\newblock On improving the selection of {Thellier}‐type paleointensity data.
\newblock {\em Geochem. Geophys. Geosyst.}, 15(4):1180--1192.

\bibitem[Pravdivtseva et~al., 2013]{pravdivtseva_i-xe_2013}
Pravdivtseva, O., Meshik, A., and Hohenberg, C.~M. (2013).
\newblock The {I}-{Xe} record: early onset of aqueous alteration in magnetites
  separated from {CM} and {CV} chondrites.
\newblock {\em Lunar Planet Sci. Conf. XLIV}, page abstract \#3104.

\bibitem[Scally and Clarke, 2001]{scally_destruction_2001}
Scally, A. and Clarke, C. (2001).
\newblock Destruction of protoplanetary discs in the {Orion} {Nebula}
  {Cluster}.
\newblock {\em Monthly Notices of the Royal Astronomical Society},
  325(2):449--456.

\bibitem[Schrader et~al., 2021]{schrader_fes_2021}
Schrader, D.~L., Davidson, J., McCoy, T.~J., Zega, T.~J., Russell, S.~S.,
  Domanik, K.~J., and King, A.~J. (2021).
\newblock The {Fe}/{S} ratio of pyrrhotite group sulfides in chondrites: {An}
  indicator of oxidation and implications for return samples from asteroids
  {Ryugu} and {Bennu}.
\newblock {\em Geochimica et Cosmochimica Acta}, 303:66--91.

\bibitem[Scott et~al., 1992]{scott_shock_1992}
Scott, E. R.~D., Keil, K., and Stöffler, D. (1992).
\newblock Shock metamorphism of carbonaceous chondrites.
\newblock {\em Geochimica et Cosmochimica Acta}, 56(12):4281--4293.

\bibitem[Scott et~al., 2018]{scott_isotopic_2018}
Scott, E. R.~D., Krot, A.~N., and Sanders, I.~S. (2018).
\newblock Isotopic dichotomy among meteorites and its bearing on the
  protoplanetary disk.
\newblock {\em The Astrophysical Journal}, 854(2):164.

\bibitem[Shadmehri and Ghoreyshi, 2019]{shadmehri_time-dependent_2019}
Shadmehri, M. and Ghoreyshi, S.~M. (2019).
\newblock Time-dependent evolution of the protoplanetary discs with magnetic
  winds.
\newblock {\em Monthly Notices of the Royal Astronomical Society},
  488(4):4623--4637.

\bibitem[Smith, 2009]{smith_use_2009}
Smith, R.~J. (2009).
\newblock Use and misuse of the reduced major axis for line-fitting.
\newblock {\em American Journal of Physical Anthropology}, 140(3):476--486.
\newblock \_eprint: https://onlinelibrary.wiley.com/doi/pdf/10.1002/ajpa.21090.

\bibitem[Sridhar et~al., 2021]{sridhar_constraints_2021}
Sridhar, S., Bryson, J. F.~J., King, A.~J., and Harrison, R.~J. (2021).
\newblock Constraints on the ice composition of carbonaceous chondrites from
  their magnetic mineralogy.
\newblock {\em Earth and Planetary Science Letters}, 576:117243.

\bibitem[Stephenson, 1993]{stephenson_three-axis_1993}
Stephenson, A. (1993).
\newblock Three-axis static alternating field demagnetization of rocks and the
  identification of natural remanent magnetization, gyroremanent magnetization,
  and anisotropy.
\newblock {\em Journal of Geophysical Research: Solid Earth}, 98(B1):373--381.
\newblock \_eprint:
  https://agupubs.onlinelibrary.wiley.com/doi/pdf/10.1029/92JB01849.

\bibitem[Sterenborg and Crowley, 2013]{sterenborg_thermal_2013}
Sterenborg, M.~G. and Crowley, J.~W. (2013).
\newblock Thermal evolution of early solar system planetesimals and the
  possibility of sustained dynamos.
\newblock {\em Phys. Earth Planet. Inter.}, 214:53--73.

\bibitem[Sutton et~al., 2017]{sutton_bulk_2017}
Sutton, S., Alexander, C. M.~O., Bryant, A., Lanzirotti, A., Newville, M., and
  Cloutis, E.~A. (2017).
\newblock The bulk valence state of {Fe} and the origin of water in chondrites.
\newblock {\em Geochimica et Cosmochimica Acta}, 211:115--132.

\bibitem[Suzuki et~al., 2016]{suzuki_evolution_2016}
Suzuki, T.~K., Ogihara, M., Morbidelli, A., Crida, A., and Guillot, T. (2016).
\newblock Evolution of protoplanetary discs with magnetically driven disc
  winds.
\newblock {\em Astronomy \& Astrophysics}, 596:A74.
\newblock Publisher: EDP Sciences.

\bibitem[Tauxe and Staudigel, 2004]{tauxe_strength_2004}
Tauxe, L. and Staudigel, H. (2004).
\newblock Strength of the geomagnetic field in the {Cretaceous} {Normal}
  {Superchron}: {New} data from submarine basaltic glass of the {Troodos}
  {Ophiolite}.
\newblock {\em Geochemistry, Geophysics, Geosystems}, 5(2).
\newblock \_eprint:
  https://agupubs.onlinelibrary.wiley.com/doi/pdf/10.1029/2003GC000635.

\bibitem[Tikoo et~al., 2012]{tikoo_magnetic_2012}
Tikoo, S.~M., Weiss, B.~P., Buz, J., Lima, E.~A., Shea, E.~K., Melo, G., and
  Grove, T.~L. (2012).
\newblock Magnetic fidelity of lunar samples and implications for an ancient
  core dynamo.
\newblock {\em Earth and Planetary Science Letters}, 337-338:93--103.

\bibitem[Tikoo et~al., 2014]{tikoo_decline_2014}
Tikoo, S.~M., Weiss, B.~P., Cassata, W.~S., Shuster, D.~L., Gattacceca, J.,
  Lima, E.~A., Suavet, C., Nimmo, F., and Fuller, M.~D. (2014).
\newblock Decline of the lunar core dynamo.
\newblock {\em Earth and Planetary Science Letters}, 404:89--97.

\bibitem[Tikoo et~al., 2017]{tikoo_two-billion-year_2017}
Tikoo, S.~M., Weiss, B.~P., Shuster, D.~L., Suavet, C., Wang, H., and Grove,
  T.~L. (2017).
\newblock A two-billion-year history for the lunar dynamo.
\newblock {\em Science Advances}.
\newblock Publisher: American Association for the Advancement of Science.

\bibitem[Torrano et~al., 2021]{torrano_relationship_2021}
Torrano, Z.~A., Schrader, D.~L., Davidson, J., Greenwood, R.~C., Dunlap, D.~R.,
  and Wadhwa, M. (2021).
\newblock The relationship between {CM} and {CO} chondrites: {Insights} from
  combined analyses of titanium, chromium, and oxygen isotopes in {CM}, {CO},
  and ungrouped chondrites.
\newblock {\em Geochimica et Cosmochimica Acta}, 301:70--90.

\bibitem[Wang et~al., 2017]{wang_lifetime_2017}
Wang, H., Weiss, B.~P., Bai, X.-N., Downey, B.~G., Wang, J., Wang, J., Suavet,
  C., Fu, R.~R., and Zucolotto, M.~E. (2017).
\newblock Lifetime of the solar nebula constrained by meteorite paleomagnetism.
\newblock {\em Science}, 355(6325):623--627.

\bibitem[Weiss et~al., 2021]{weiss_history_2021}
Weiss, B.~P., Bai, X.-N., and Fu, R.~R. (2021).
\newblock History of the solar nebula from meteorite paleomagnetism.
\newblock {\em Science Advances}, 7(1):eaba5967.
\newblock Publisher: American Association for the Advancement of Science
  Section: Review.

\bibitem[Weiss and Bottke, 2021]{weiss_what_2021}
Weiss, B.~P. and Bottke, W.~F. (2021).
\newblock What can meteorites tell us about the formation of {Jupiter}?
\newblock {\em AGU Advances}, 2(2):e2020AV000376.
\newblock \_eprint:
  https://agupubs.onlinelibrary.wiley.com/doi/pdf/10.1029/2020AV000376.

\bibitem[Weiss et~al., 2010]{weiss_paleomagnetic_2010}
Weiss, B.~P., Gattacceca, J., Stanley, S., Rochette, P., and Christensen, U.~R.
  (2010).
\newblock Paleomagnetic records of meteorites and early planetesimal
  differentiation.
\newblock {\em Space Science Reviews}, 152(1):341--390.

\bibitem[Weiss and Tikoo, 2014]{weiss_lunar_2014}
Weiss, B.~P. and Tikoo, S.~M. (2014).
\newblock The lunar dynamo.
\newblock {\em Science}, 346(6214).
\newblock Publisher: American Association for the Advancement of Science
  Section: Review.

\bibitem[Weiss et~al., 2017]{weiss_nonmagnetic_2017}
Weiss, B.~P., Wang, H., Sharp, T.~G., Gattacceca, J., Shuster, D.~L., Downey,
  B., Hu, J., Fu, R.~R., Kuan, A.~T., Suavet, C., Irving, A.~J., Wang, J., and
  Wang, J. (2017).
\newblock A nonmagnetic differentiated early planetary body.
\newblock {\em Earth Planet. Sci. Lett.}, 468:119--132.

\bibitem[Yu et~al., 2004]{yu_toward_2004}
Yu, Y., Tauxe, L., and Genevey, A. (2004).
\newblock Toward an optimal geomagnetic field intensity determination
  technique.
\newblock {\em Geochem. Geophys. Geosyst.}, 5:doi:10.1029/2003GC000630.

\bibitem[Zhu et~al., 2011]{zhu_transitional_2011}
Zhu, Z., Nelson, R.~P., Hartmann, L., Espaillat, C., and Calvet, N. (2011).
\newblock Transitional and pre-transitional disks: gap opening by multiple
  planets?
\newblock {\em The Astrophysical Journal}, 729(1):47.
\newblock Publisher: IOP Publishing.

\end{thebibliography}
\endgroup

\end{document}